\documentclass[a4paper,fleqn,usenatbib]{mnras}

\usepackage[T1]{fontenc}
\usepackage{ae,aecompl}

\usepackage{graphicx}   % Including figure files
\usepackage{amsmath}    % Advanced maths commands
\usepackage{amssymb}    % Extra maths symbols

\newcommand{\be}{\begin{equation}}

\newcommand{\ee}{\end{equation}}
\newcommand{\bea}{\begin{eqnarray}}
\newcommand{\eea}{\end{eqnarray}}
\newcommand{\bef}{\begin{figure}}
\newcommand{\eef}{\end{figure}}
\newcommand{\bce}{\begin{center}}
\newcommand{\ece}{\end{center}}
\def\lsim{\mathrel{\rlap{\lower4pt\hbox{\hskip1pt$\sim$}}
    \raise1pt\hbox{$<$}}}         %less than or approx. symbol
\def\gsim{\mathrel{\rlap{\lower4pt\hbox{\hskip1pt$\sim$}}
    \raise1pt\hbox{$>$}}}         %greater than or approx. symbol
\title[Fast spinning strange stars]{Fast spinning strange stars: possible ways to constrain interacting quark matter parameters}

\author[Bhattacharyya et al.]{Sudip Bhattacharyya$^{1}$\thanks{E-mail: sudip@tifr.res.in},
	Ignazio Bombaci$^{2,3,4}$, Domenico Logoteta$^{3}$, Arun V. Thampan$^{5,6}$\\
	$^{1}$Department of Astronomy and Astrophysics, Tata Institute of Fundamental Research, Mumbai 400005, India\\
    $^{2}$Dipartimento di Fisica, Universit\`{a} di Pisa, Largo B. Pontecorvo, 3 I-56127 Pisa, Italy\\ 
    $^{3}$INFN, Sezione di Pisa, Largo B. Pontecorvo, 3 I-56127 Pisa, Italy\\ 
    $^{4}$European Gravitational Observatory, Via E. Amaldi, I-56021 S. Stefano a Macerata, Cascina Italy\\ 
    $^{5}$Department of Physics, St. Joseph's College, 36 Lalbagh Road, Bangalore 560027, India\\
    $^{6}$Inter-University Centre for Astronomy and Astrophysics (IUCAA), India}

\date{Accepted XXX. Received YYY; in original form ZZZ}

\pubyear{2015}

\begin{document}
\label{firstpage}
\pagerange{\pageref{firstpage}--\pageref{lastpage}}
\maketitle

% Abstract of the paper
\begin{abstract}
For a set of equation of state (EoS) models involving interacting strange quark matter,
characterized by an effective bag constant ($B_{\rm eff}$) and a perturbative QCD corrections 
term ($a_4$), we construct fully general relativistic 
equilibrium sequences of rapidly spinning strange stars
for the first time. Computation of such sequences is important to study millisecond pulsars
and other fast spinning compact stars. Our EoS models can support
a gravitational mass ($M_{\rm G}$) and a spin frequency ($\nu$) at least
up to $\approx 3.0 M_\odot$ and $\approx 1250$ Hz respectively, 
and hence are fully consistent with measured $M_{\rm G}$
and $\nu$ values. This paper reports the effects of 
$B_{\rm eff}$ and $a_4$ on measurable compact star properties, which could be useful to 
find possible ways to constrain these fundamental quark matter parameters,
within the ambit of our EoS models.
We confirm that a lower $B_{\rm eff}$ allows a higher mass. Besides, for known $M_{\rm G}$ and
$\nu$, measurable parameters, such as stellar radius, radius-to-mass ratio and 
moment of inertia, increase with the decrease of $B_{\rm eff}$.
Our calculations also show that $a_4$ significantly affects the stellar rest
mass and the total stellar binding energy. As a result, $a_4$ can have signatures 
in evolutions of both accreting and non-accreting compact stars, and the observed 
distribution of stellar mass and spin and other source parameters.
Finally, we compute the parameter values of two important pulsars, PSR J1614-2230 and 
PSR J1748-2446ad, which may have implications to probe their evolutionary 
histories, and for constraining EoS models.
\end{abstract}


%%%%%%%%%%%%%%%%%%%%%%%%%%%%%%%%%%%%%%%%%%%%%%%%%%

%%%%%%%%%%%%%%%%% BODY OF PAPER %%%%%%%%%%%%%%%%%%

\section{Introduction}\label{Introduction}

Despite more than 80 years since the proposal of neutron stars as a new class of 
astrophysical objects \citep{BZ34} and about 50 years since their first discovery 
as radio pulsars \citep{Hew+68}, 
their true nature and internal composition still remain one of the most 
fascinating enigma in modern astrophysics.   

The bulk properties and the internal constitution of compact stars (neutron stars) primarily 
depend on the equation of state (EoS) of strong interacting matter, 
i.e. on the thermodynamical relation between 
the matter pressure, energy density and temperature. Determining the correct EoS model
describing the interior of compact stars is a fundamental problem of physics, and a major effort has been made
during the last few decades to solve this problem by measuring the stellar bulk properties.
A number of these measurement methods for low-mass X-ray binaries (LMXBs), 
based on thermonuclear X-ray bursts, regular X-ray pulsations, high-frequency 
quasi-periodic oscillations, broad relativistic spectral emission lines, quiescent
emissions and orbital motions, have been reviewed in \citet{Bhattacharyya2010}. Spectral
and timing properties of non-accreting millisecond radio pulsars can also be useful 
to measure stellar mass and radius (e.g., \citet{Bogdanov2009,Bogdanov2008}). However,
until recently theoretically proposed EoS models could not be effectively constrained because of
systematic uncertainties (e.g., \citet{Bhattacharyya2010}).

The recent precise measurement of the mass ($1.97\pm0.04 M_\odot$) of the 
millisecond pulsar PSR J1614-2230 has ruled out all EoS models which cannot support
such high values of masses %$\gsim 2.0 M_\odot$ 
\citep{demo10}. However, in response to this discovery, 
new realistic EoS models,
that support high mass, have been proposed. So essentially all types of EoS models, such as
nucleonic, hyperonic, strange quark matter, hybrid, still survive, and it is still a very
important problem to constrain compact star EoS models.

In this paper, we are interested in fast spinning compact stars.
So far the spin frequencies of a number of such stars have been measured
(e.g., \citet{Watts2012,PatrunoWatts2012,Smedleyetal2014}). Some of these sources are binary 
millisecond radio pulsars, and the masses of a fraction of them have been relatively
precisely measured (e.g., \citet{Smedleyetal2014}). In order to constrain EoS models, three
independent bulk parameters of a given compact star are to be measured, and the third 
observable parameter (after the mass and spin) could be the stellar radius \citep{Loetal2013,Bogdanov2009} 
or the moment of inertia \citep{Morrisonetal2004}. Here we note that a very high observed 
spin frequency could constrain EoS models, but so far the highest spin frequency
observed is 716 Hz from a radio pulsar PSR J1748-2446ad \citep{Hessels2006}. 
This spin frequency is allowed by almost all proposed EoS models, and hence 
this spin measurement alone is not a useful property to constrain these models.

From a fundamental point of view, the EoS of strongly interacting matter should be derived 
by numerically solving quantum chromodynamics (QCD) equations on a space-time lattice (lattice QCD).  
Since the central density of compact stars can significantly exceed the saturation density  
($\sim 2.8 \times 10^{14}$~g/cm$^{3}$) of nuclear matter and their temperature could be considered 
equal to zero after a few minutes of their formation \citep{BurLat86,bomb+95,prak97}, 
these compact stars can be viewed as natural laboratories to explore the phase diagram of QCD
in the low temperature $T$ and high baryon chemical potential  $\mu_b$ region.
In this region of the QCD phase diagram a transition to a phase with deconfined quarks and gluons 
is expected to occur and to influence a number of interesting astrophysical phenomena 
\citep{pg10,sot11,r1,blv07,r6,weis11,nish12}.

Recent high precision lattice QCD calculations at zero baryon chemical potential
(i.e. zero baryon density) have clearly shown that at high temperature and for physical values of the quark 
masses, quarks and gluons become the most relevant degrees of freedom. 
The transition to this quark gluon plasma phase is a crossover \citep{bern05,cheng06,aoki06} rather 
than a real phase transition.  
Lattice QCD calculations in this regime have also distinctly demonstrated the importance of taking 
into account the interactions of quarks and gluons since the calculated EoS significantly 
deviates from that of an ideal gas. 

Unfortunately, current lattice QCD calculations at finite baryon chemical potential are 
plagued with the so called ``sign problem'', which makes them unrealizable by all 
known lattice methods.      
Thus, to explore the QCD phase diagram at low temperature T and high $\mu_b$, it is 
necessary to invoke some approximations in QCD or to apply some QCD effective model. 

Along these lines, for example, a model of the equation of state (EoS) 
of strange quark matter (SQM) \citep{mit-eos} inspired by the MIT bag model of hadrons \citep{mit} 
has been extensively used by many authors to calculate the structure of 
strange stars \citep{witt84,afo86,hzs86,XDLi99a,XDLi99b,Xu99}, or the structure of the 
so called hybrid stars, {\it i.e.} compact stars with an SQM core.   
In this model SQM is treated as an ideal relativistic Fermi gas of
{\it up} ({\it u}), {\it down} ({\it d}) and {\it strange} ({\it s}) quarks  
(together with an appropriate number of electrons to guarantee electric charge neutrality and 
equilibrium with respect to the weak interactions), 
that reside in a region characterized by a constant energy density $B$. 
The parameter $B$ takes into account, in a crude phenomenological 
manner,  the nonperturbative aspects of QCD and is related to the bag constant which 
in the MIT bag model \citep{mit} gives the confinement of quarks within hadrons.   

The deconfinement phase transition has been also described using an EoS of quark 
gluon plasma derived within the Field Correlator Method (FCM) \citep{dosh87,DS88,simo88,digiac02}  
extended to finite baryon chemical potential \citep{ST07a,ST07b,sim1,sim3,sim5}. 
FCM  is a nonperturbative approach to QCD which includes from first principles, 
the dynamics of confinement.  
The model is parametrized in terms of the gluon condensate $G_2$ and the large distance static 
quark-antiquark ($Q\bar{Q}$) potential $V_1$.  
These two quantities control the EoS of the denconfined phase at fixed quark masses and temperature. 
The main constructive characteristic of FCM is the possibility to describe the whole 
QCD phase diagram  as it can span from high temperature and low baryon chemical potential,  
to low $T$ and high $\mu_{\rm b}$ limit. 

Recently, \citet{bom-log13} have established that the values of gluon condensate $G_2$ 
extracted from the measured mass $M = 1.97 \pm 0.04 \, M_\odot$ of PSR~J1614-2230 \citep{demo10}
are fully consistent with the values of the same quantity derived  within FCM, from 
lattice QCD calculations of the deconfinement transition temperature at zero baryon chemical 
potential \citep{bors10,baz12}.  
FCM thus provides a powerful tool to link numerical calculations of QCD on 
a space-time lattice with measured compact star masses \citep{logbomb13}.   

In this paper we make use of a more traditional approach to compute the EoS of SQM.  
In fact, the simple version of the MIT bag model EoS can be extended to include perturbative 
corrections due to quark interactions, up to the second order ($\mathcal{O}(\alpha_{\rm s}^2)$) in 
the strong structure constant $\alpha_s$ \citep{FMc77,FMc78,Bal78,Fra01,Kur10} 
{\footnote{In \citet{mit-eos} the EoS for strange quark matter was calculated 
up to the order $\mathcal{O}(\alpha_{\rm s})$.}}. 
Within this modified bag model one can thus evaluate the non-ideal behaviour of the EoS 
of cold SQM at high density.  

The modified bag model EoS has already been used to calculate the structure of non-spinning 
strange stars \citep{Fra01,Alf05,weis11,Fra14} and hybrid stars \citep{Alf05,weis11}. 

As already mentioned, there are two types of compact stars containing SQM. 
The first type is represented by the so called hybrid stars, i.e. compact stars containing a 
quark-hadron mixed core and eventually a pure SQM inner core. The second type, strange stars, 
is realized  when SQM satisfies the Bodmer--Witten hypothesis \citep{bod71,witt84}. 
According to this hypothesis SQM is absolutely stable, in other words, its energy per baryon $(E/A)_{uds}$ 
is less than that of the most bound atomic nuclei ($^{56}\rm{Fe}, ^{58}\rm{Fe}, ^{62}\rm{Ni}$) which 
is $\sim 930.4~\rm{MeV}$.  
The absolute stability of SQM does not preclude the existence of ``ordinary" matter \citep{bod71,witt84}. 
In fact, under this hypothesis, atomic nuclei can be considered as metastable states (with respect 
to the decay to SQM droplets) having a mean-life time which is many orders of magnitude larger 
than the age of the Universe.  

In the last few decades, many researchers (see e.g. \citet{xu01,web05} and references therein) 
have tried to identify possible clear observational signatures to distinguish whether a compact star 
is a strange star, a hybrid star or a ``normal" neutron star (nucleonic star).

The mass-radius (M-R) relation is one of the most promising compact star features to solve this puzzle. 
In fact, ``low mass" (i.e. $M \lesssim 1\, M_\odot$) strange stars have $M \sim R^3$, whereas 
normal neutron stars, in the mass range between $0.5~M_\odot$ and $\sim 0.7~M_{max}$ (where $M_{max}$ is 
the stellar sequence maximum mass), have a radius which is almost independent on the mass \citep{lp01,bomb07}
This qualitative difference in the M-R relation is due to the fact that strange stars are self-bound objects    
whereas normal neutron stars are bound by gravity. 
Constraints for the M-R relation, extracted from observational data of compact stars in the X-ray sources 
SAX~J1808.4-3658 \citep{XDLi99a} and  4U~1728-34 \citep{XDLi99b} seem to indicate that these objects  
could be accreting strange stars. Note that the observations of thermonuclear X-ray bursts, 
which are believed to originate from unstable thermonuclear burning of accumulated 
accreted matter on the compact star surface, cannot
preclude these LMXBs from having strange stars. This is because, according to a 
proposed model \citep{StejnerMadsen2006}, the bursts could happen on a thin crust of ``ordinary" matter  
(i.e. a solid layer consisting of a Coulomb lattice of atomic nuclei in 
$\beta$-equilibrium with a relativistic electron gas, similar to the outer crust of a neutron star) 
separated from the main body of the strange star by a strong Coulomb 
barrier \citep{afo86,StejnerMadsen2005}. 

Other significant observational informations on the constitution of compact stars may come out 
in the next few years from the expected detection of gravitational waves from compact stars with 
ground-based interferometers, such as Advanced VIRGO and Advanced LIGO. 
In fact, it has been shown by different research groups, e.g. \citep{ander02,benhar07,fu08,ander11,rup13},   
that gravitational waves driven by r-mode instabilities or from f-, p- and g-mode oscillations 
could be able to discriminate among different types of compact stars. 

Another interesting possibility is that both ``normal" neutron stars (nucleonic stars) and 
quark stars (i.e. strange stars or hybrid stars) could exist in nature \citep{r0,r1,bo04}. 
In this scenario quark stars could be formed via a conversion process of metastable normal neutron 
stars \citep{bd00,r0,r1}. 

Several experimental searches for strangelets (small lumps of absolutely stable SQM) have been undertaken using different techniques. 
For example the Alpha Magnetic Spectrometer (AMS-02) \citep{AMS}, on board of the International Space Station 
since May 2011, could be able to detect strangelets in cosmic rays with excellent charge resolution up 
to an atomic number $Z \sim 26$. 
Strangelets search in lunar soil, brought back by the NASA Apollo 11 mission, have been performed using 
the tandem accelerator at the  Wright Nuclear Structure Laboratory at Yale \citep{han09}
 
It is important to emphasize that all the present observational data and our present experimental and 
theoretical knowledge of the properties of dense matter, do not allow us to accept or to exclude 
the validity of the Bodmer--Witten hypothesis. 

In this paper, we assume the validity of the Bodmer--Witten hypothesis and 
we compute equilibrium sequences of rapidly spinning strange stars in general relativity.  
In \S~\ref{EoS}, we describe the modified bag model EoS \citep{Fra01,Alf05,weis11} 
for SQM used in our calculations and the two parameters, effective bag constant ($B_{\rm eff}$) 
and perturbative QCD corrections term parameter ($a_4$), which characterize this model.
In \S~\ref{Computation}, we discuss the method to compute parameters of 
fast spinning compact stars in stable configurations. Here we also describe various
limit sequences.
In \S~\ref{Results}, we present the numbers from our numerical calculations using
tables and figures. In \S~\ref{Discussion}, we discuss the implications of our results,
especially for constraining $B_{\rm eff}$ and $a_4$ using observations.
In \S~\ref{Summary}, we summarize the key points of this paper.

%%%%%%%%%%%%%
\section{The modified bag model equation of state for strange quark matter}\label{EoS} 
%%%%%%%%%%%%%

The EoS for strange quark matter including the effects of gluon mediated QCD interactions between 
quarks up to $\mathcal{O}(\alpha_{\rm s}^2)$ 
can be written in a straightforward and easy-to-use form similar to the simple and popular version of the 
MIT bag model EoS. The grand canonical potential per unit volume takes the form 
(we use units where $\hbar = 1$, and $c = 1$)   
\be
\label{eos}
\Omega = \sum_{i=u,d,s,e} \Omega_i^0 +  \frac{3}{4\pi^2}(1 - a_4)\Big(\frac{\mu_{\rm b}}{3}\Big)^4 + B_{\rm eff}\, .
\ee
where $\Omega_i^0$ is the grand canonical potential for {\it u}, {\it d}, {\it s} quarks 
and electrons described as ideal relativistic Fermi gases. 
We take $m_u = m_d = 0$, $m_s = 100~\rm{MeV}$ and $m_e = 0$.  
The second term on the right hand side of Eq.(\ref{eos}) 
accounts for the perturbative QCD corrections to $\mathcal{O}(\alpha_s^2)$ 
\citep{Fra01,Alf05,weis11} and its value represents the degree of deviations from  
an ideal relativistic Fermi gas  EoS,
with $a_4 = 1$ corresponding to the ideal gas. 
The baryon chemical potential $\mu_b$ can be written in terms of the {\it u}, {\it d} and {\it s} 
quark chemical potentials as $\mu_b = \mu_u + \mu_d + \mu_s$.  
The term $B_{\rm eff}$ is an effective bag constant which takes into accounts in a phenomenological 
way of nonperturbative aspects of QCD.  

Using standard thermodynamical relations, the energy density can be written as: 
\be
\label{endens}
      \varepsilon =  \sum_{i=u,d,s,e} \Omega_i^0 + \frac{3}{4\pi^2}(1 - a_4) \Big(\frac{\mu_b}{3}\Big)^4 
                    + \sum_{i=u,d,s,e}{\mu_i n_i} + B_{eff} \, ,
\ee
where $n_i$ is the number density for each particle species which can be calculated as 
\be
\label{numdens}
      n_i = - \bigg(\frac{\partial\Omega}{\partial \mu_i}\bigg)_V  
\ee
and the total baryon number density is 
\be
\label{n_B}
       n_b =  \frac{1}{3}(n_u + n_d + n_s)\, .
\ee
Equilibrium with respect to the weak interactions  implies the following relations 
between the quarks and electron chemical potentials: 
\be
   \mu_s = \mu_d  = \mu_u + \mu_e \, ,
\ee
the electric charge neutrality condition requires:
\be
    \frac{2}{3} n_u - \frac{1}{3} n_d - \frac{1}{3} n_s - n_e = 0 \, .
\ee

Since in the present paper we study the case of spinning strange stars, 
we consider values of the EoS parameters $a_4$ and  $B_{\rm eff}$ so that SQM satisfies the 
Bodmer--Witten hypothesis. 
Next, to guarantee the observed stability of atomic nuclei with respect to a possible decay to a droplet of 
non-strange (i.e. {\it u}, {\it d}) quark matter, we require that the energy per baryon $(E/A)_{ud}$ 
of non-strange quark matter should satisfy the condition  
$ (E/A)_{ud} >  930.4~\rm{MeV} + \Delta$, 
where $\Delta \sim 4~\rm{MeV}$ accounts for finite size effects of the energy per baryon of a 
droplet of non-starnge quark matter with respect to the bulk ($A  \rightarrow \infty$) case \citep{mit-eos}.   

The values for the EoS parameters considered in the calculations reported in this work, are listed in 
Table~\ref{table_EoS}. In particular, to explore the effects of the perturbative QCD corrections on the properties of spinning strange stars, for the fixed value $B_{\rm eff}^{1/4} = 138~\rm{MeV}$, we 
consider two different values of the parameter $a_4$ ($=0.61$ and $0.80$);  $a_4 = 0.61$ gives a larger deviation from the ideal gas than $a_4 = 0.80$.  

\begin{figure*}
\centering
\hspace{-0.8cm}
\includegraphics*[width=15cm]{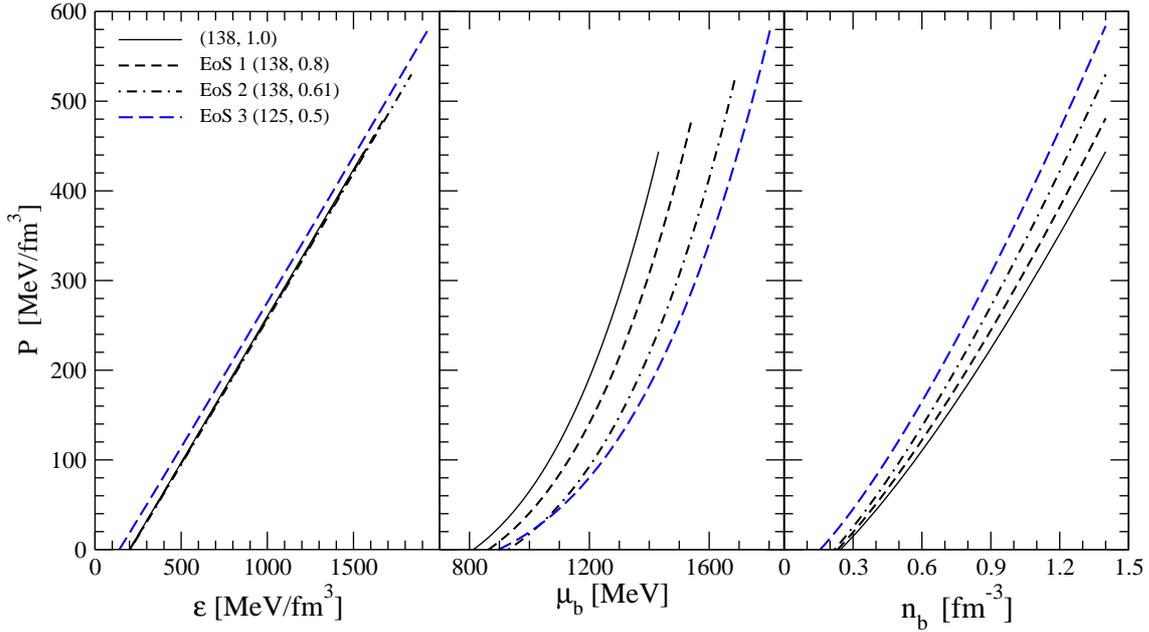}
\caption{Pressure $P$ versus energy density $\varepsilon$ (left panel), 
$P$ versus baryon chemical potential $\mu_b$ (middle panel) and 
$P$ versus baryon number density $n_b$ (right panel) 
for the three strange quark matter EoS parameter sets used in the present work 
(see \S~\ref{EoS} and Table~\ref{table_EoS}).    
To highlight the influence of the perturbative QCD corrections term $a_4$ on the EoS,  
in comparison with the EoS 1 and EoS 2 cases, we also plot the EoS with 
$B_{\rm eff}^{1/4} = 138~\rm{MeV}$ and $a_4 = 1.0$ (ideal relativistic Fermi gas plus bag pressure).
\label{eos_1.ps}}
\end{figure*}

The EoS, for the three parametrizations used in the present work, are plotted in Fig.~\ref{eos_1.ps}.  
Notice that at fixed $B_{\rm eff}$, the value of $a_4$ has a very small effect on the EoS expressed as 
$P = P(\varepsilon)$, i.e. pressure $P$ as a function of the energy density $\varepsilon$ 
(left panel in Fig.~\ref{eos_1.ps}), which enters in the structure equations for non-spinning 
and spinning stars.  
To highlight the influence of the perturbative QCD corrections term $a_4$ on the EoS,  
in comparison with the EoS 1 and EoS 2, we also plot in Fig.~\ref{eos_1.ps} that for the case 
$B_{\rm eff}^{1/4} = 138~\rm{MeV}$ and $a_4 = 1.0$ (ideal relativistic Fermi gas plus bag pressure). 
We do not, however, use this EoS in the calculations for spinning strange stars reported here, 
since we have verified that for this choice of the parameters 
$B_{\rm eff}^{1/4}$ and $a_4$, the atomic nuclei are unstable 
with respect to decay to a droplet of non-strange quark matter. 

%%%%%%%%%%%%%%%%%%%%%%%%%%%%%%
%%%%%%%%%%%%%%%%%%%%%%%%%%%%%%%%%%%%%%
    
The results plotted in the left panel of Fig.~\ref{eos_1.ps}, however, do not imply that 
the perturbative QCD corrections are unimportant.  
Clearly, looking at Eq.(\ref{eos}) and the results plotted in the middle panel of Fig.~\ref{eos_1.ps},   
one sees that the EoS in the form $P(\mu_b) = -\Omega(\mu_b)$ has a sizeable dependence 
on the parameter $a_4$. 

The influence of the parameter $a_4$ on the EoS (Eqs.(\ref{eos}), (\ref{endens})) of SQM 
is more clear in the case of massless quarks. 
In fact, in this case one can show that the EoS for SQM, in a parametrical form in terms of 
the baryon number density, can be written as:    
\begin{eqnarray}
  \varepsilon = K n_b^{4/3} +  B_{\rm eff} \nonumber \\
          P = \frac{1}{3} K n_b^{4/3} -  B_{\rm eff} \, ,
\label{eos-0}
\end{eqnarray} 
where 
\begin{eqnarray}
    K = \frac{9}{4} \frac{\pi^{2/3}}{a_4^{1/3}}  
\label{kappa}
\end{eqnarray}
and eliminating the baryon number density $n_b$ one gets 
\begin{eqnarray}
\label{eq:eosBag}
     P = \frac{1}{3} (\varepsilon - 4 B_{\rm eff}) \, ,
\end{eqnarray}
which does not depend on $a_4$. Thus the stellar gravitational mass $M_G$ versus the central 
energy density $\varepsilon_{\rm c}$, or the stellar radius $R$ versus $\varepsilon_{\rm c}$, 
in the case of massless quarks, will not depend on the perturbative QCD corrections term $a_4$. 
These stellar properties will have a tiny dependence on $a_4$ in our case due to the 
finite value of the strange quark mass, $m_s = 100~\rm{MeV}$.  
Nevertheless, stellar properties like the total rest mass of the star $M_0(\varepsilon_{\rm c})$, 
the relation between $M_G$ and $M_0$ and consequently the value of the total stellar binding 
energy $B = M_0 - M_G$ \citep{bd00} will significantly be affected by the parameter $a_4$. 
This  ultimately should affect the energetics of explosive phenomena like supernovae or 
the conversion of normal neutron stars (nucleonic stars) to strange or hybrid stars \citep{bd00,r1}.  

%%%%%%  
To illustrate how the parameter $a_4$ affects the stellar rest mass $M_0$ and binding energy $B$, 
let us consider for the moment the case of non-spinning configurations. 
The stellar rest mass is given by 
\be
\label{Mbar}
 M_0  = m_u \int_{0}^{R} 4 \pi r^2 n_{b}(r) \bigg[1 - \frac{2G m(r)}{c^2 r} \bigg]^{-1/2} dr
\ee
where $m_u = 931.494~{\rm Mev/c^2}$ is the atomic mass unit, $n_{b}(r)$ is the baryon number density at the radial coordinate $r$, and $m(r)$ is the stellar gravitational mass enclosed within $r$. 
Now, in addition, consider the case of massless quarks. 
Using Eq.(\ref{eos-0}) and (\ref{kappa}), we get the for the baryon number density 
\be
   n_{b}(r) = \bigg(\frac{2}{3}\bigg)^{3/2}\frac{1}{\sqrt{\pi}} \, a_4^{1/4} 
              \Big[\varepsilon(r) - B_{eff}\Big]^{3/4} \, .
\label{nbar}
\ee
Substituting this expression in Eq. (\ref{Mbar}) and considering that the functions 
$m(r)$ and $\varepsilon(r)$  do not depend on the parameter $a_4$,   
one obtains the following scaling relation for the stellar rest mass  
\be
   M_0(\varepsilon_c; a_4) = \bigg(\frac{a_4}{a_4^{\prime}}\bigg)^{1/4} M_0(\varepsilon_c; a_4^{\prime}) 
\label{scal_M0}
\ee
where $a_4$ and $a_4^{\prime}$ are two different values of the perturbative QCD corrections parameter 
and $\varepsilon_c$ is the central energy density of the star. 
Therefore, for a star with a given gravitational mass $M_G$, the stellar binding energy scales with the parameter $a_4$ as
\be
   B(a_4) = \bigg(\frac{a_4}{a_4^{\prime}}\bigg)^{1/4} B(a_4^{\prime}) \,.
\label{scal_M0}
\ee

%%%%%%%%%%%%%%%%%%%%%%%%%%%%%%
\begin{figure*}
\centering
\hspace{-0.8cm}
\includegraphics*[width=15cm]{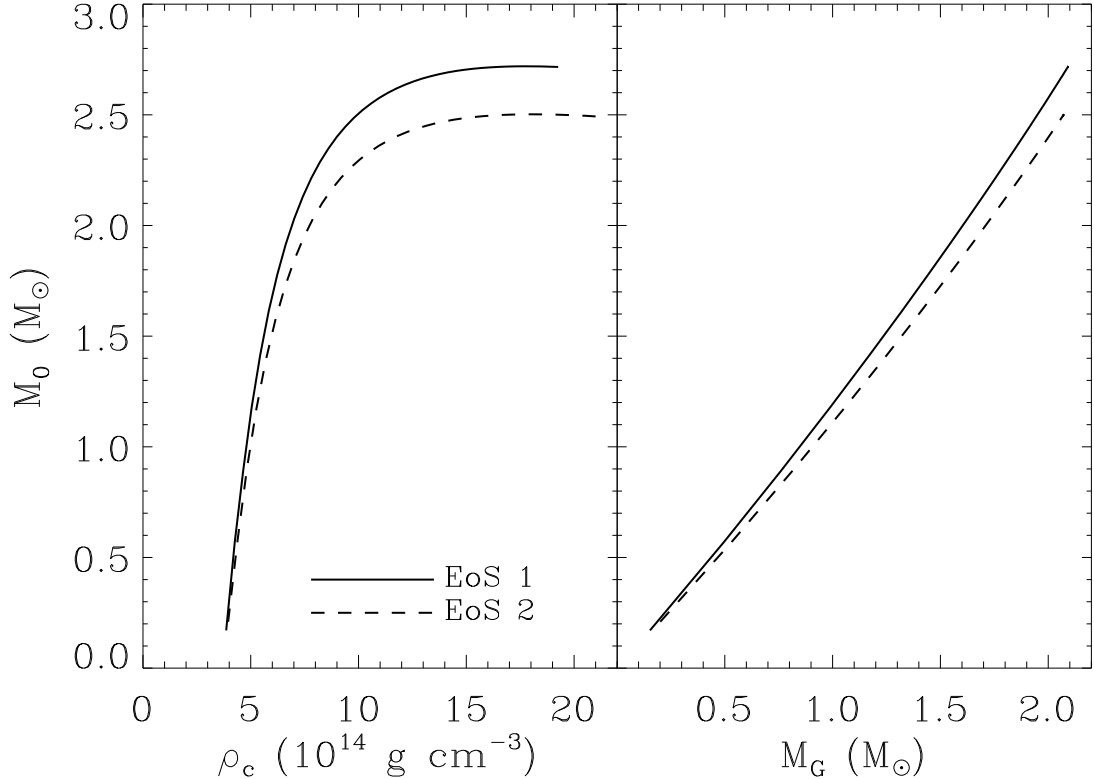} 
\caption{Stellar rest mass $M_0$ versus central density (left panel), and 
$M_0$ versus stellar gravitational mass $M_G$ (right panel) for non-spinning strange stars.   
\label{Mg-Mb}}
\end{figure*}
%%%%%%%%%%%%%%%%%%%%%%%%%%%%%%%%%%%%%%

In Fig.~\ref{Mg-Mb} we report our results for the rest mass $M_0$ of the star as a function 
of its central density (left panel) and as a function of the corresponding gravitational mass $M_G$ 
(right panel). Results in Fig.~\ref{Mg-Mb} are regarding non-spinning strange stars with the 
EoS~1 ($a_4 = 0.80$) and EoS~2 ($a_4 = 0.61$) with $B_{\rm eff}^{1/4} = 138~\rm{MeV}$ in both cases. 
As we can see, decreasing the value of the parameter $a_4$, i.e. increasing the deviation from 
the ideal relativistic Fermi gas ($a_4 = 1$), results in a reduction of the stellar rest mass 
at a given central density or at a given $M_G$. Notice that the results plotted in Fig.~\ref{Mg-Mb}, 
for a strange quark mass $m_s = 100~\rm{MeV}$, are well reproduced by the scaling relation 
given in Eq.~(\ref{scal_M0}). 
As we will see in the following, this scaling relation will be also very useful in the case 
of fast spinning strange stars. 

Simple scaling relations for the properties of non-spinning strange stars have been obtained 
in terms of the effective bag constant $B_{eff}$ \citep{witt84,hzs86,bomb99, Haenseletal2007} 
in the case of the EoS given by Eq.~(\ref{eq:eosBag}).  
As discussed in Appendix~\ref{appendix}, the gravitational mass and the radius of the star,  
for two different values $B_{\rm eff,1}$ and $B_{\rm eff,2}$ of the effective bag constant, 
are related by 
\be
 M_G(\varepsilon_{\rm c,1};B_{\rm eff,1}) = 
                  \bigg(\frac{B_{\rm eff,2}}{B_{\rm eff,1}}\bigg)^{1/2} M_G(\varepsilon_{\rm c,2};B_{\rm eff,2}), 
\label{eq:scal5aa}
\ee
\be
R(\varepsilon_{\rm c,1};B_{\rm eff,1}) = 
                  \bigg(\frac{B_{\rm eff,2}}{B_{\rm eff,1}}\bigg)^{1/2} R(\varepsilon_{\rm c,2};B_{\rm eff,2}), 
\label{eq:scal5bb}
\ee
with the two central energy densities satisfying the condition
\be
\varepsilon_{\rm c,1}/\varepsilon_{\rm c,2} = B_{\rm eff,1}/B_{\rm eff,2}.
\label{eq:rho6b}
\ee
Equations~(\ref{eq:scal5aa}) and (\ref{eq:scal5bb}) give the scaling law for the mass-radius relation. 
In particular they hold \citep{witt84,hzs86} for the maximum mass configuration.

%%%%%
Finally, the perturbative QCD corrections to the EoS make it possible to fulfil the Bodmer--Witten 
hypothesis in a region of the $B_{\rm eff}^{1/4}$--$a_4$ plane where one can get strange stars with a maximum gravitational 
mass significantly larger than $2~M_\odot$ (see Fig. 1 in \citet{weis11}), 
and thus in agreement with current measurements of compact star masses like that of 
PSR~J1614-2230 ($M = 1.97 \pm 0.04 \, M_\odot$; \citet{demo10}).

%%%%%%%%%%%%%%%%%%%%%%%%%%%%%%%%%%%%%%%%%%%%%%%%%%%%%%%%%%%%%%%%%%%%%%%%%%%%%%%%
\section{Computation of rapidly spinning stellar structures}\label{Computation}

For computing the stationary and equilibrium sequences of strange stars, 
we make use of the formalism
mentioned in \citet{Cooketal1994, Dattaetal1998, Bombacietal2000, Bhattacharyyaetal2000, BhattacharyyaBhattacharyaThampan2001, 
 BhattacharyyaMisraThampan2001, BhattacharyyaThampanBombaci2001, Bhattacharyya2002, Bhattacharyya2011}. The
general spacetime for such a star is (using $c=G=1$; \citet{Bardeen1970, Cooketal1994}):
\begin{eqnarray}
{\rm d}s^2 = -{\rm e}^{\gamma+\rho}{\rm d}t^2 + {\rm e}^{2\alpha}({\rm d}r^2 + 
r^2{\rm d}\theta^2) + {\rm e}^{\gamma-\rho}r^2\sin^2\theta \nonumber \\
({\rm d}\phi - \omega{\rm d}t)^2,
\end{eqnarray}
where $\gamma$, $\rho$, $\alpha$ are metric potentials, $\omega$
is the angular speed of the stellar fluid relative to the local inertial frame,
and $t$, $r$ and $\theta$ are temporal, quasi-isotropic radial and polar angular
coordinates respectively. 
Assuming the matter comprising the star to be a perfect fluid with energy momentum tensor:
\begin{eqnarray}
    T^{\mu \nu} = (\varepsilon + P) u^{\mu} u^{\nu} + P g^{\mu \nu}
\end{eqnarray}
and a unit time-like description for the four velocity vector we can decompose
the Einstein field equations projected onto the frame of reference of a Zero Angular Momentum Observer (ZAMO)
to yield three elliptic equations for $\gamma$, $\rho$ and $\omega$ and a
linear ordinary differential equation for $\alpha$ \citep{Cooketal1994}. The elliptic equations are
then converted to integral equations using the Green's function approach.
The hydrostatic equations are derived from the relativistic equations assuming a linear spin
law (in our case the spin law integral vanishes for the rigid spin condition).  The 
solution for the hydrostatic equilibrium equation reduces to solving of one algebraic equation
at each grid position and matching these with the equatorial values for self-consistency.
For a desired central density $\rho_{\rm c}$ and polar radius to equatorial radius ratio ($r_{\rm p}/r_{\rm e}$),
a multi-iteration run of the program to achieve self-consistency yields a 2-dimensional distribution of the 
metric potentials, density and pressure repesenting an equilibrium solution.  This equilibrium solution is then 
used to compute compact star parameters, such as gravitational mass ($M_{\rm G}$), rest
mass ($M_{\rm 0}$), equatorial circumferential radius ($R$),
total angular momentum ($J$), spin frequency ($\nu$), moment of inertia ($I$),
total spinning kinetic energy ($T$), total gravitational energy ($W$),
surface polar redshift ($Z_{\rm p}$), and forward ($Z_{\rm f}$) and 
backward ($Z_{\rm b}$) redshifts for tangential emission of photons at
the equator \citep{Cooketal1994, Dattaetal1998}.

Once the equilibrium parameters describing the structure are obtained, it becomes 
feasible to compute general quantities exterior to the compact star like the Keplerian
angular speed and specifically, the radius $r_{\rm ISCO}$ of the innermost stable circular orbit (ISCO).
These quantities depend on the effective potential that has a maximum at $r_{\rm ISCO}$. 
Here is how $r_{\rm ISCO}$ is calculated.
The radial equation of motion around such a compact star is 
\.{r}$^2 \equiv {\rm e}^{2\alpha + \gamma + \rho}({\rm d}r/{\rm d}\tau)^2 = $\~{E}$^2 - $\~{V}$^2$, 
where, ${\rm d}\tau$ is the proper time, \~{E} is the specific
energy, which is a constant of motion, and \~{V} is the effective potential.
This effective potential is given by \~{V}$^2 = {\rm e}^{\gamma+\rho}[1 + 
\dfrac{l^2/r^2}{{\rm e}^{\gamma-\rho}}] + 2\omega$\~{E}$l - \omega^2 l^2$,
where $l$ is the specific angular momentum and a constant of motion.
$r_{\rm ISCO}$ is determined using the condition \~{V}$_{,rr}$ = 0,
where a comma followed by one $r$ represents a first-order
partial derivative with respect to $r$ and so on \citep{ThampanDatta1998}. We consider 
$r_{\rm orb}$ ($= r_{\rm ISCO})$ to be the smallest possible radius of the accretion disc.
But $R$ is the absolute lower limit of the disc radius.
Therefore, if the star extends beyond the ISCO,
we set $r_{\rm orb} = R$ \citep{Bhattacharyyaetal2000, Bhattacharyya2011}.

Since observations of some pulsars have provided a measure for $M_{\rm G}$ and $\nu$, 
we compute constant $M_{\rm G}$ and constant $\nu$ equilibrium sequences of a couple of pulsars. 
For each EoS model, we also compute a number of constant $M_{\rm 0}$ sequences.
These sequences would represent the evolution of isolated compact stars conserving their rest mass. 
Such sequences are stable to quasi-radial mode perturbations if 
$\dfrac{\partial J}{\partial \rho_{\rm c}}|_{M_{\rm 0}}$ $< 0$; we calculate the limit where this 
inequality does not hold.
In addition to this instability limit, there are three other limits:
(1) the static or nonspinning limit, where $\nu \rightarrow 0$
and $J \rightarrow 0$; (2) the mass-shed limit, at which the compact star
spins too fast to keep matter bound to the surface; and (3) the low-mass limit, 
below which a compact star cannot form.
These four limits together define the stable stellar parameter space for 
an EoS model \citep{Cooketal1994}. Here, apart from the instability limit,
we calculate the static and mass-shed limit sequences, 
but we do not attempt to determine the low-mass limit, where the numerical
solutions are less accurate \citep{Cooketal1994}.

\section{Results}\label{Results}

\begin{figure}
\centering
\hspace{-1.0cm}
\includegraphics*[width=9cm]{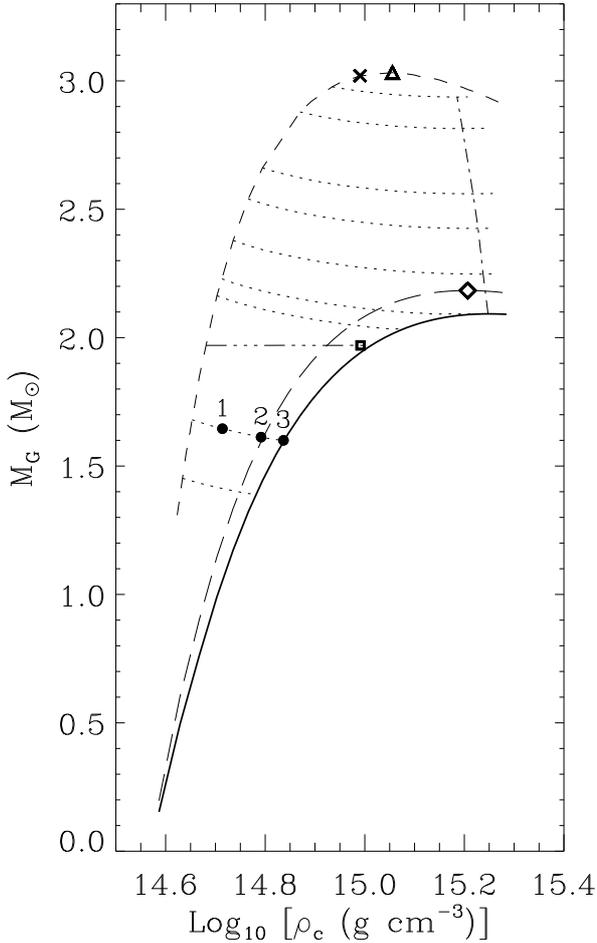}
\caption{Gravitational mass versus central density plot of strange stars for
EoS 1 (see \S~\ref{EoS} and Fig.~\ref{eos_1.ps}). The solid curve is
for a non-spinning star, the short-dashed curve shows the mass-shed limit,
the long-dashed curve is for a spin frequency of 716 Hz, the dotted curves
are for evolutionary (i.e., constant rest mass) sequences, the dash-dot
curve gives the instability limit to quasi-radial mode perturbations, and
the dash-triple-dot curve is for the gravitational mass = 1.97 M$_\odot$. 
The constant rest mass values, for dotted curves from 
bottom to top, are 1.71 M$_\odot$, 2.00 M$_\odot$, 2.63 M$_\odot$, 2.72 M$_\odot$,
2.92 M$_\odot$, 3.15 M$_\odot$, 3.33 M$_\odot$, 3.66 M$_\odot$ and 3.82 M$_\odot$.
The triangle and cross symbols are for the maximum mass (Table~\ref{table_mass_shed_max_mass}) 
and the maximum total angular momentum (Table~\ref{table_mass_shed_max_ang_mom}) 
respectively for the mass-shed sequence, the three filled circles marked with
`1', `2' and `3' on the rest mass sequence $M_{\rm 0} = 2.00$ M$_\odot$ correspond to the `No.'
column of Table~\ref{rest_mass_seq},
the square symbol is for the observed mass (= 1.97 M$_\odot$) and spin
frequency (= 317.5 Hz) of PSR J1614-2230 (Table~\ref{table_1.97}), and
the diamond symbol is for the maximum mass which can be supported 
by the fastest known pulsar PSR J1748-2446ad
spinning at 716 Hz (Table~\ref{table_716}). See \S~\ref{Results} for a description.
\label{mgvslogd1388.ps}}
\end{figure}

\begin{figure}
\centering
\hspace{-1.0cm}
\includegraphics*[width=9cm]{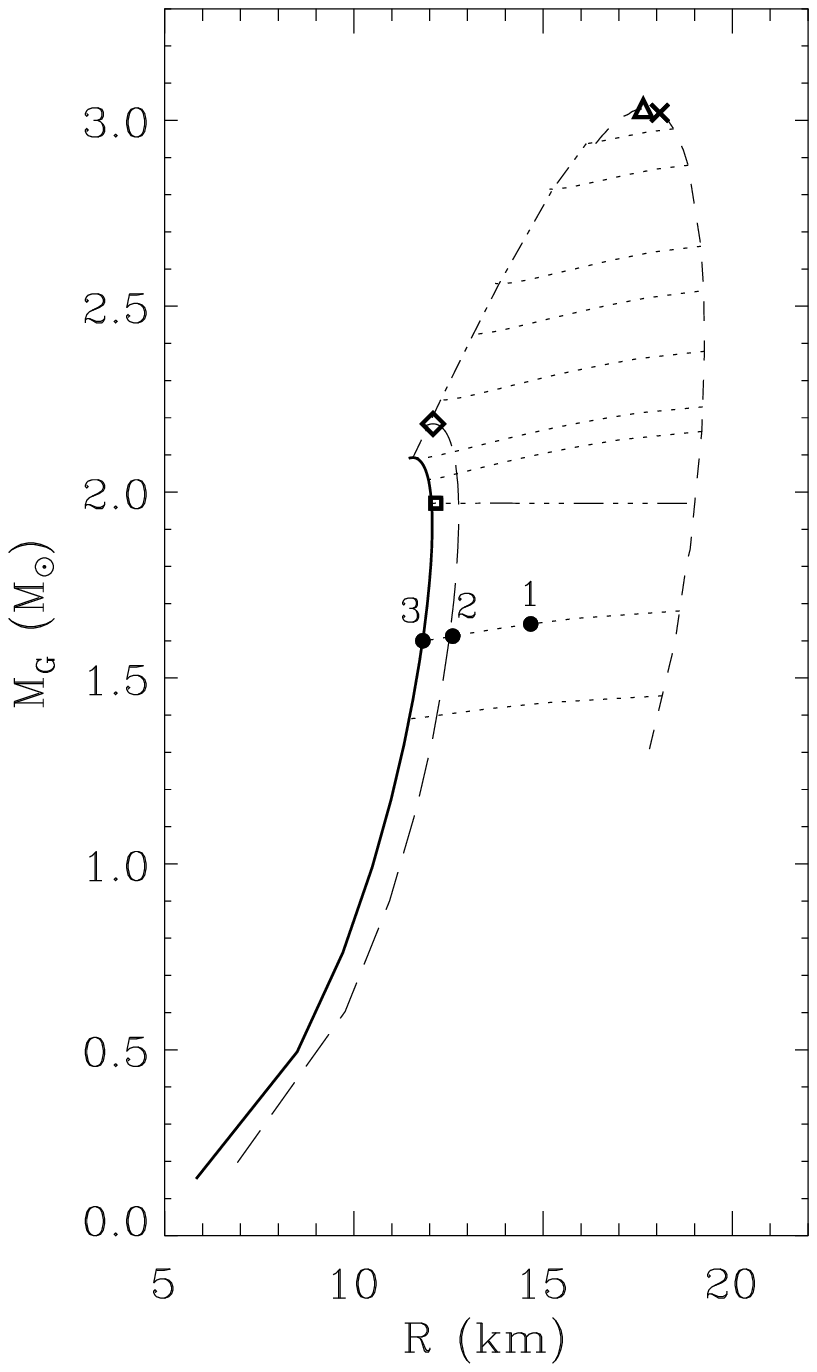}
\caption{Gravitational mass versus equatorial radius plot of strange stars
for EoS 1 (see \S~\ref{EoS} and Fig.~\ref{eos_1.ps}).
The meanings of curves and symbols are same as in Fig.~\ref{mgvslogd1388.ps}.
\label{mgvsre1388.ps}}
\end{figure}

\begin{figure}
\centering
\hspace{-1.0cm}
\includegraphics*[width=9cm]{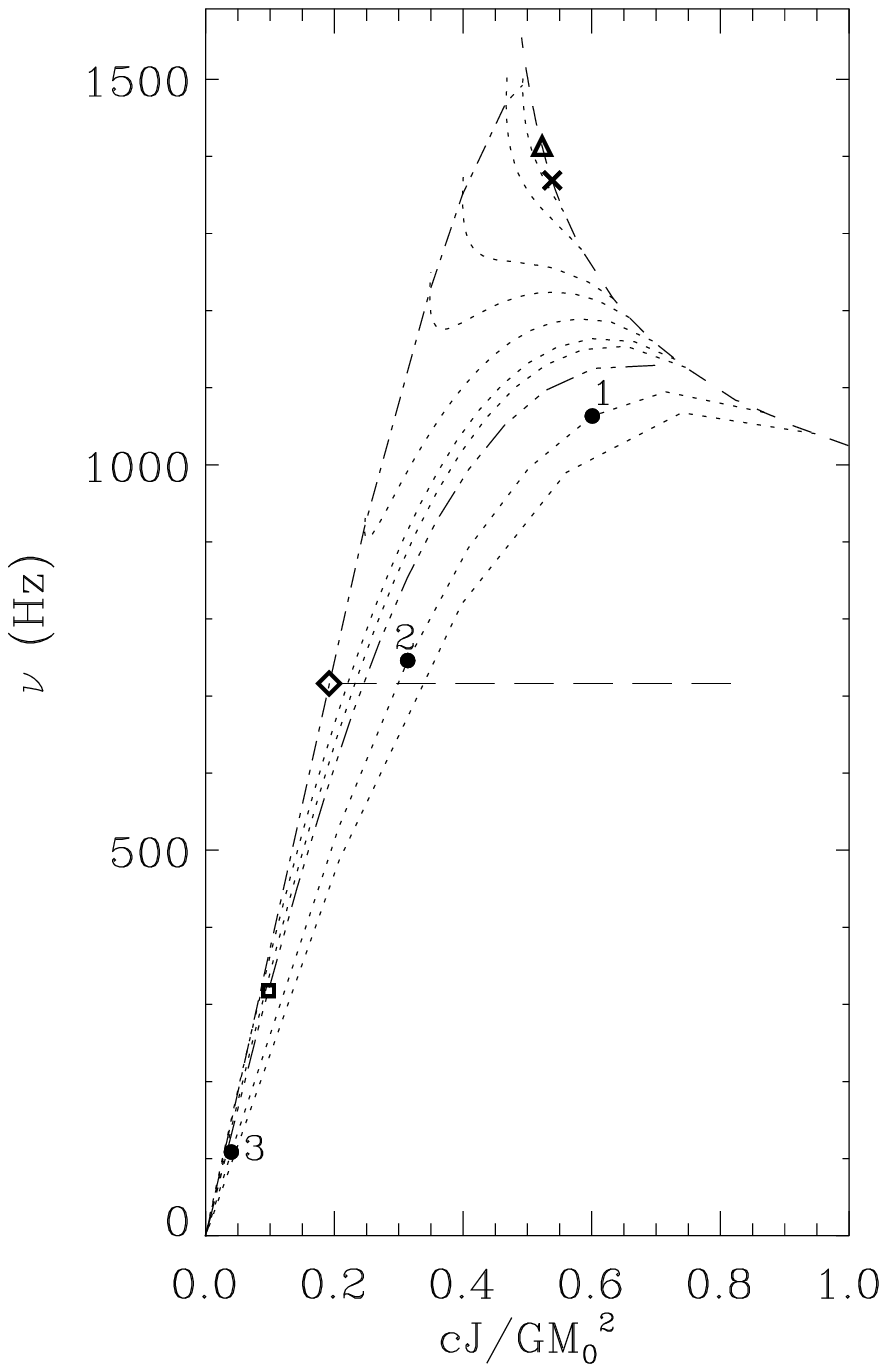}
\caption{Spin frequency versus dimensionless angular momentum plot of strange stars
for EoS 1 (see \S~\ref{EoS} and Fig.~\ref{eos_1.ps}).
$J$ is the total angular momentum and $M_0$ is the rest mass.
The meanings of curves and symbols are same as in Fig.~\ref{mgvslogd1388.ps}.
\label{omvsj1388.ps}}
\end{figure}

\begin{figure}
\centering
\hspace{-1.0cm}
\includegraphics*[width=9cm]{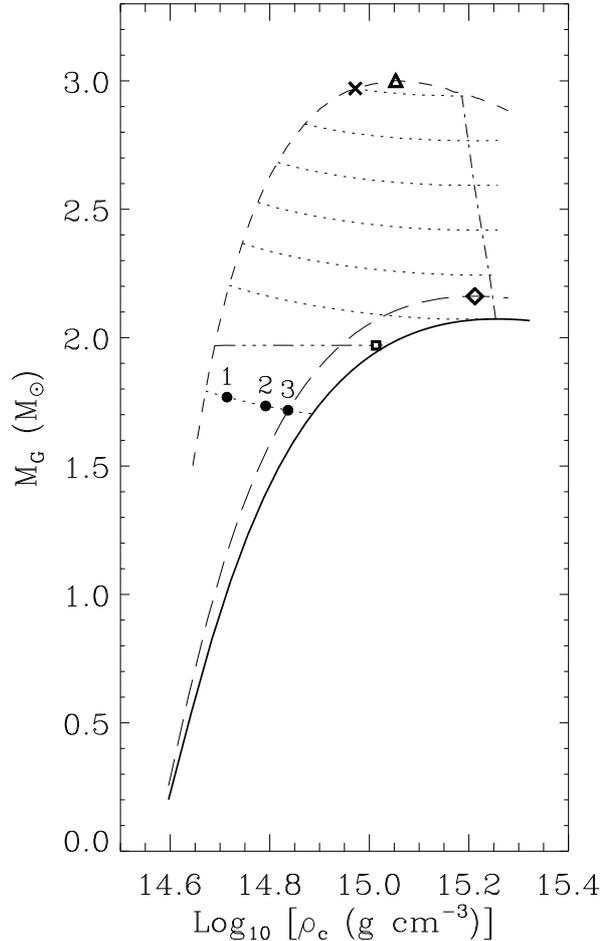}
\caption{Gravitational mass versus central density plot of strange stars for
EoS 2 (see \S~\ref{EoS} and Fig.~\ref{eos_1.ps}). 
The meanings of curves and symbols are same as in Fig.~\ref{mgvslogd1388.ps}.
The constant rest mass values, for dotted curves from 
bottom to top, are 2.00 M$_\odot$, 2.50 M$_\odot$, 2.71 M$_\odot$, 2.92 M$_\odot$,
3.13 M$_\odot$, 3.34 M$_\odot$, and 3.55 M$_\odot$.
\label{mgvslogd1386.ps}}
\end{figure}

\begin{figure}
\centering
\hspace{-1.0cm}
\includegraphics*[width=9cm]{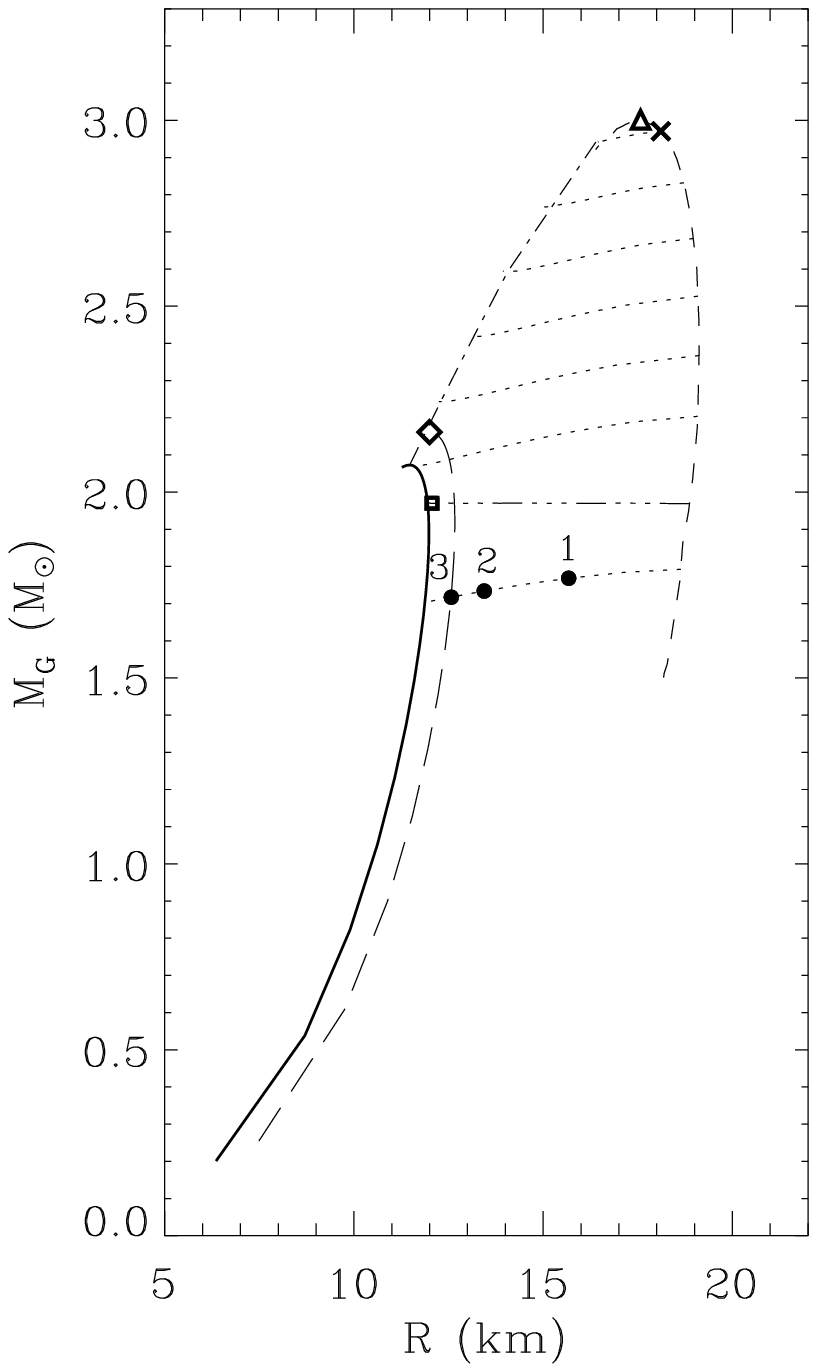}
\caption{Gravitational mass versus equatorial radius plot of strange stars for
EoS 2 (see \S~\ref{EoS} and Fig.~\ref{eos_1.ps}). 
The meanings of curves and symbols are same as in Fig.~\ref{mgvslogd1388.ps}.
\label{mgvsre1386.ps}}
\end{figure}

\begin{figure}
\centering
\hspace{-1.0cm}
\includegraphics*[width=9cm]{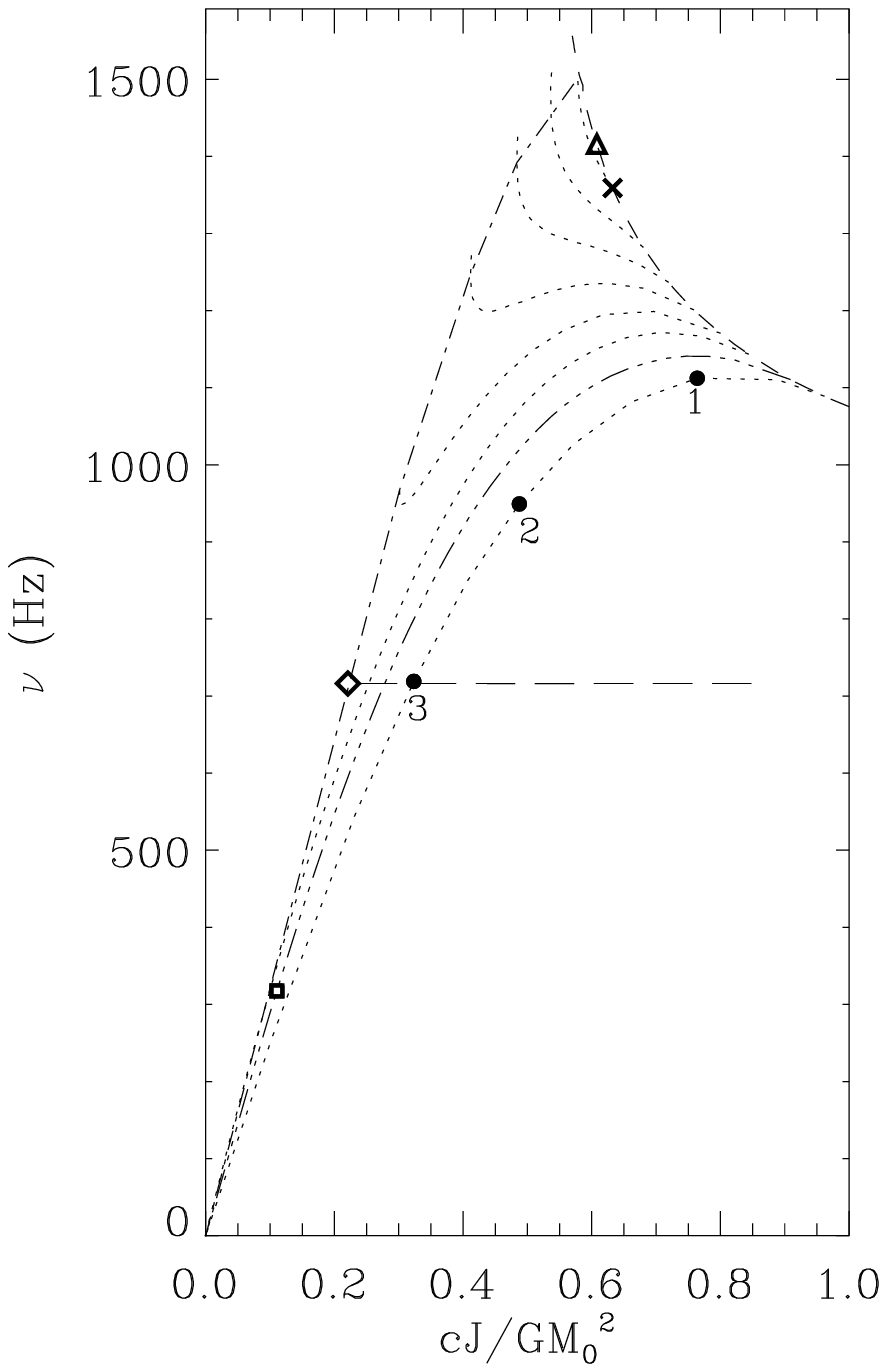}
\caption{Spin frequency versus dimensionless angular momentum plot of strange stars for
EoS 2 (see \S~\ref{EoS} and Fig.~\ref{eos_1.ps}). 
$J$ is the total angular momentum and $M_0$ is the rest mass.
The meanings of curves and symbols are same as in Fig.~\ref{mgvslogd1388.ps}.
\label{omvsj1386.ps}}
\end{figure}

\begin{figure}
\centering
\hspace{-1.0cm}
\includegraphics*[width=9cm]{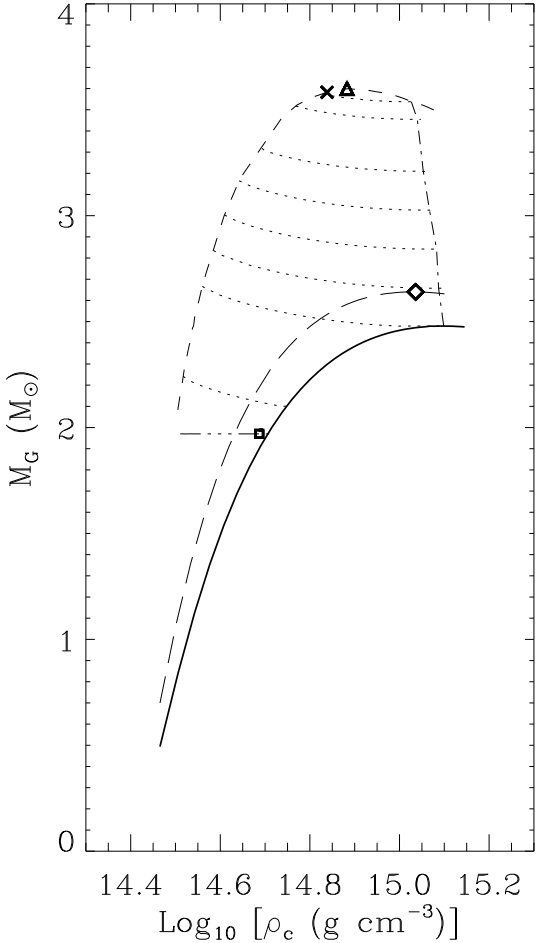}
\caption{Gravitational mass versus central density plot of strange stars for
EoS 3 (see \S~\ref{EoS} and Fig.~\ref{eos_1.ps}). 
The meanings of curves and symbols are same as in Fig.~\ref{mgvslogd1388.ps}.
The constant rest mass values, for dotted curves from 
bottom to top, are 2.54 M$_\odot$, 3.09 M$_\odot$, 3.32 M$_\odot$, 3.54 M$_\odot$,
3.77 M$_\odot$, 4.00 M$_\odot$, 4.30 M$_\odot$, and 4.40 M$_\odot$.
\label{mgvslogd1255.ps}}
\end{figure}

\begin{figure}
\centering
\hspace{-1.0cm}
\includegraphics*[width=9cm]{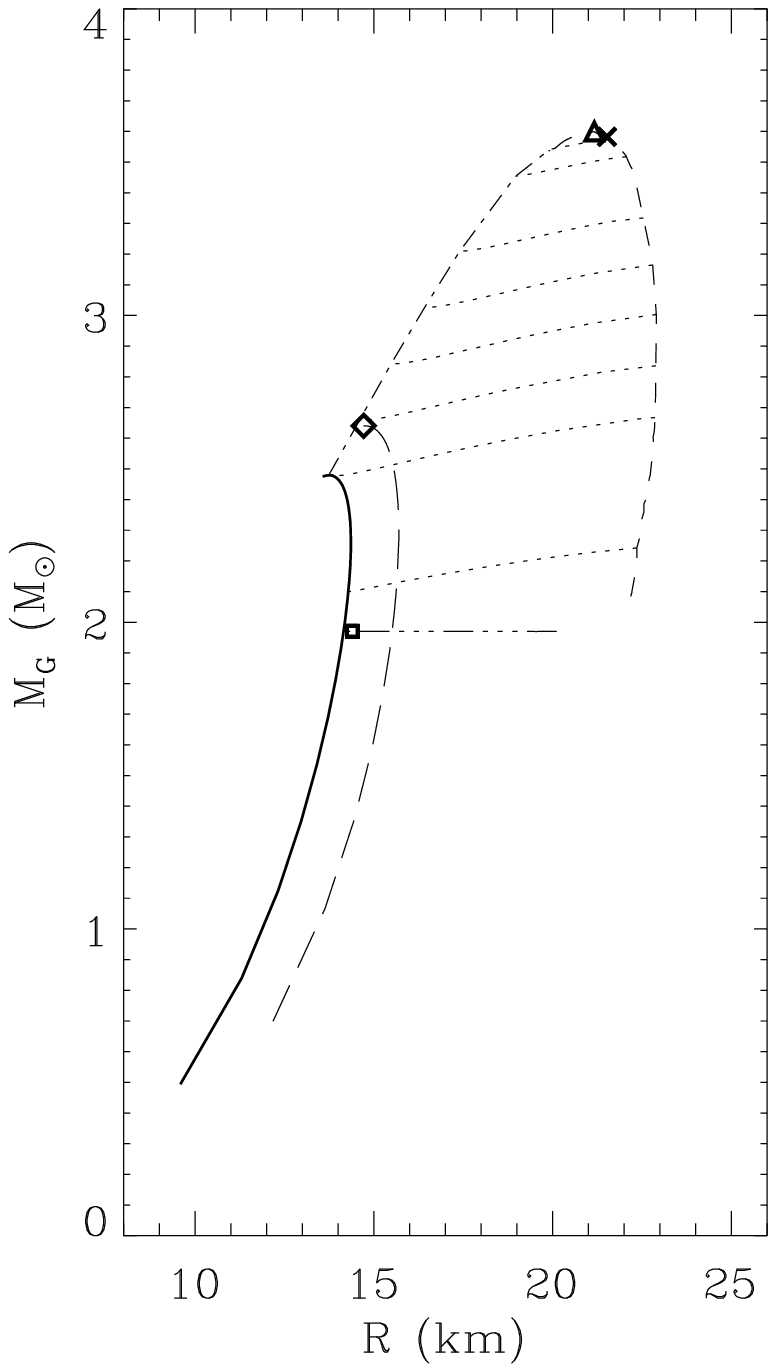}
\caption{Gravitational mass versus equatorial radius plot of strange stars for
EoS 3 (see \S~\ref{EoS} and Fig.~\ref{eos_1.ps}). 
The meanings of curves and symbols are same as in Fig.~\ref{mgvslogd1388.ps}.
\label{mgvsre1255.ps}}
\end{figure}

\begin{figure}
\centering
\hspace{-1.0cm}
\includegraphics*[width=9cm]{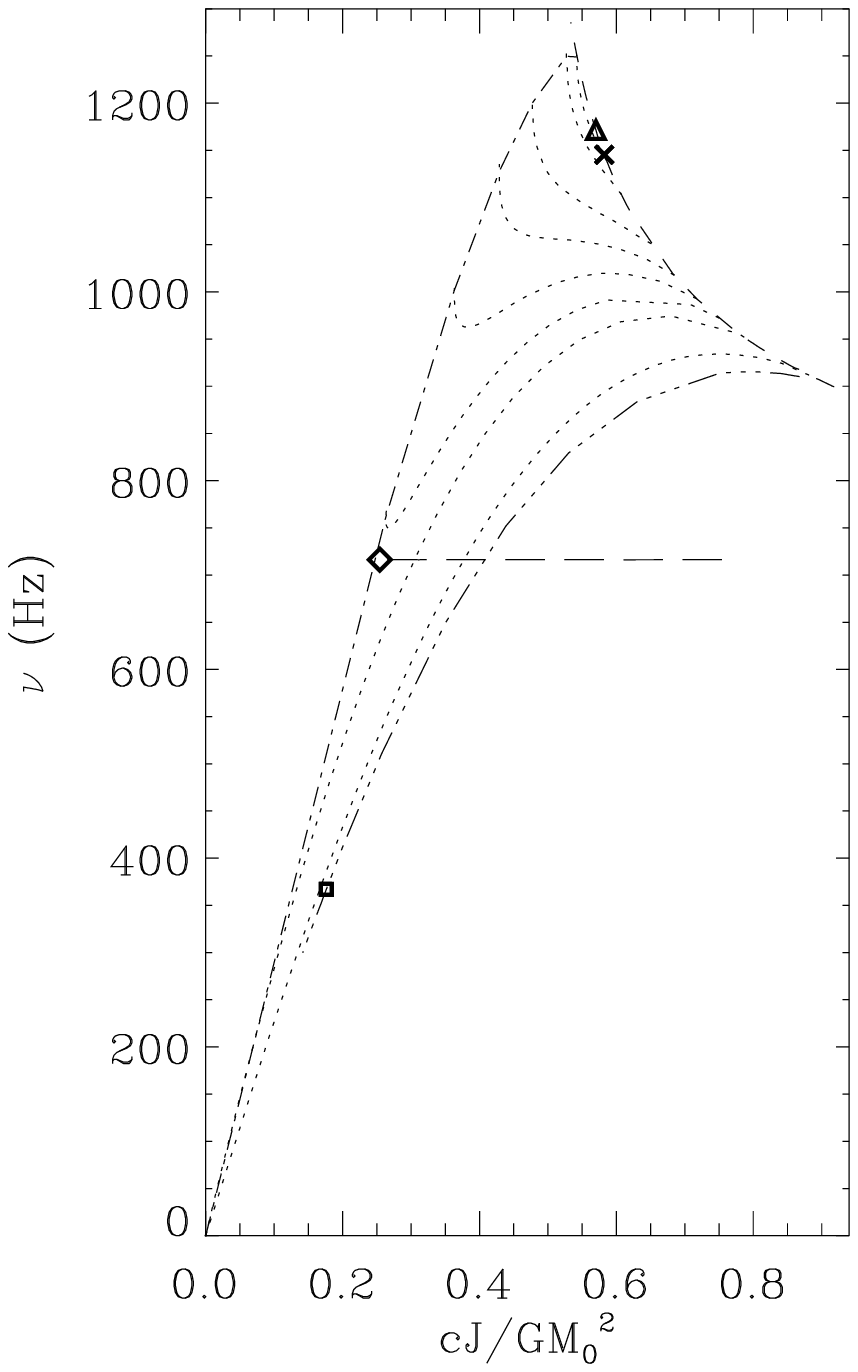}
\caption{Spin frequency versus dimensionless angular momentum plot of strange stars for
EoS 3 (see \S~\ref{EoS} and Fig.~\ref{eos_1.ps}). 
$J$ is the total angular momentum and $M_0$ is the rest mass.
The meanings of curves and symbols are same as in Fig.~\ref{mgvslogd1388.ps}.
\label{omvsj1255.ps}}
\end{figure}

The results of our computations are summarized in Figs.~\ref{mgvslogd1388.ps} through
\ref{omvsj1255.ps} and Tables~\ref{table_non_spinning} through \ref{table_716}.
Figs.~\ref{mgvslogd1388.ps}--\ref{omvsj1388.ps} are for EoS 1, 
Figs.~\ref{mgvslogd1386.ps}--\ref{omvsj1386.ps} are for EoS 2, and
Figs.~\ref{mgvslogd1255.ps}--\ref{omvsj1255.ps} are for EoS 3.
Figs.~\ref{mgvslogd1388.ps}, \ref{mgvslogd1386.ps} and \ref{mgvslogd1255.ps}
show $M_{\rm G}$ versus $\rho_{\rm c}$ curves.
The static limits, which can support maximum $M_{\rm G}$ values of 2.09 M$_{\odot}$,
$2.07$ M$_\odot$ and $2.48$ M$_\odot$ for EoS models 1, 2 and 3 respectively,
are shown by solid curves. As $\rho_{\rm c}$ increases, more inward gravitational pull,
i.e., more $M_{\rm G}$ is required to balance the pressure.
Apart from the static limit, the mass-shed and instability limit sequences
are also shown in these figures (see \S~\ref{Computation}). The star is not stable
for the parameter space to the right of the curve defining the instability limit.
A number of constant rest mass sequences, and
the $M_{\rm G} = 1.97$ M$_\odot$ sequence are also shown in 
Figs.~\ref{mgvslogd1388.ps}, \ref{mgvslogd1386.ps} and \ref{mgvslogd1255.ps}.
On a constant rest mass sequence, the stellar $J$ increases from right to left,
and so does $M_{\rm G}$ to balance the extra centrifugal force.
The rest mass sequence, which joins the maximum $M_{\rm G}$ on the static limit,
separates the supramassive sequence region (above) from the
normal sequence region (below). In the supramassive region, a compact star
is so massive that it can be stable only if it has sufficient $J$ value.
When $J$ decreases (say, via electromagnetic and/or gravitational radiation) 
to the value corresponding to the instability limit, the compact star collapses further
to become a black hole. Another surprising aspect of the supramassive region is,
as $J$ decreases, $I$ can decrease at a higher rate, and hence $\nu$ can increase.
So the star can spin up, as it loses angular momentum.

Figs.~\ref{mgvsre1388.ps}, \ref{mgvsre1386.ps} and \ref{mgvsre1255.ps}
show $M_{\rm G}$ versus $R$ curves for various sequences. These figures show that
$R$ usually increases with $M_{\rm G}$. This property distinguishes strange
stars from ``normal" neutron stars (see also \S~\ref{Introduction}). 
Figs.~\ref{omvsj1388.ps}, \ref{omvsj1386.ps} and \ref{omvsj1255.ps}
show $\nu$ versus dimensionless $J$ ($cJ/GM_0^2$) curves. These figures clearly
show, while $\nu$ decreases with the decrease of $J$ in the normal sequence region,
it can have opposite behaviour in the supramassive region.
Finally, we note that all the curves of the Figs.~\ref{mgvslogd1388.ps}--\ref{omvsj1255.ps}
are qualitatively consistent with those for other SQM EoS models (see, for example,
\citet{Bombacietal2000}). 

Table~\ref{table_non_spinning} displays the values of $\rho_{\rm c}$, 
$M_{\rm 0}$, $R$, $R/r_{\rm g}$ and $r_{\rm orb}$ for the maximum $M_{\rm G}$ 
configurations in static limit. This table shows that the stellar radius is 3.74$r_{\rm g}$
(3.75$r_{\rm g}$, 3.75$r_{\rm g}$) and the ISCO is $\sim 7$ ($\sim 7$, $\sim 8$) km above 
the compact star for EoS 1 (2, 3). Here, $r_{\rm g}$
is the Schwarzschild radius. 
For the mass-shed limit, the maximum values of $M_{\rm G}$, $J$ and $\nu$
are $3.03$ M$_\odot$, $7.17\times10^{49}$ g cm$^2$ s$^{-1}$ and $\sim 1500$ Hz respectively
for EoS 1, $3.00$ M$_\odot$, $7.01\times10^{49}$ g cm$^2$
s$^{-1}$ and $\sim 1501$ Hz respectively for EoS 2, and
are $3.60$ M$_\odot$, $9.98\times10^{49}$ g cm$^2$
s$^{-1}$ and $\sim 1250$ Hz respectively for EoS 3 (see Tables~\ref{table_mass_shed_max_mass} and
\ref{table_mass_shed_max_ang_mom}).
But note that these three numbers for a given EoS model are for three different configurations
on the mass-shed limit sequence. Tables~\ref{table_mass_shed_max_mass} and 
\ref{table_mass_shed_max_ang_mom}, which display the stable parameter values for 
maximum values of $M_{\rm G}$ and $J$ respectively, show that
there is a gap of almost 2 km between the star and the ISCO, and $T/W$ is
roughly 0.21 for each EoS model.

Table~\ref{rest_mass_seq} displays three configurations on the rest mass 
$M_{\rm 0} = 2.00 M_{\odot}$ sequence for EoS 1 and EoS 2. 
These configurations are also marked with filled circles in Figs.~\ref{mgvslogd1388.ps}--\ref{omvsj1386.ps}.
This table shows that, for the same $\rho_{\rm c}$, most of the parameters have significantly
different values for the two EoS models. This difference is around or more than
one order of magnitude for $J$, $\nu$ and $T/W$ for the highest $\rho_{\rm c}$ value
considered.

After this general characterisation of the SQM EoS models for rapidly
spinning stars, we compute the stable structures of two well known pulsars.
We show the parameter values for the massive pulsar PSR J1614-2230 
($M_{\rm G} = $ 1.97 M$_\odot$, $\nu = 317.5$ Hz; \S~\ref{Introduction}) 
for the three EoS models in Table~\ref{table_1.97}. 
This pulsar is marked with a square symbol in each of 
Figs.~\ref{mgvslogd1388.ps}--\ref{omvsj1255.ps}. The mass of the fastest
known pulsar PSR J1748-2446ad ($\nu = 716$ Hz; \S~\ref{Introduction}) is
not yet known. We, therefore, compute the $\nu = 716$ Hz equilibrium 
sequences for the three EoS models, and show the stable stellar parameter 
values for a number of configurations in Table~\ref{table_716}. The
maximum values of $M_{\rm G}$ supported by $\nu = 716$ Hz are
2.18 M$_\odot$, 2.16 M$_\odot$ and 2.64 M$_\odot$ for EoS models 1, 2 and 3 respectively.
Figs.~\ref{mgvslogd1388.ps}--\ref{omvsj1255.ps} show the $\nu = 716$ Hz
sequence with long-dashed curves and the maximum $M_{\rm G}$ configuration
with a diamond symbol on each of these curves.

\section{Discussion}\label{Discussion}

In this paper, we confirm that strange stars with interacting quark matter can support 
$> 2 M_\odot$ gravitational mass, even when they are not spinning. 
Since the highest precisely measured masses of compact stars are 
$\approx 2 M_\odot$ (see \S~\ref{Introduction}; also \citet{Antoniadisetal2013}), 
this provides a possibility 
of the existence of strange stars. In order to explore this possibility,
here we consider three EoS models based on MIT bag model with perturbative corrections due to
quark interactions (\S~\ref{EoS}). These EoS models are characterized by an effective 
bag constant ($B_{\rm eff}$) and a perturbative QCD corrections term ($a_4$).
We, for the first time, compute rapidly spinning strange
star equilibrium sequences for these models, which are essential to study millisecond
pulsars, as well as non-pulsar fast spinning compact stars in LMXB systems (\S~\ref{Introduction}).
We find that the maximum masses supported by our EoS models are in the range 
$3.0-3.6 M_\odot$ for the mass-shed limit, and in the range $\approx 2.2-2.6 M_\odot$
for the spin frequency ($\nu = 716$ Hz) of the fastest known pulsar. Hence, a precise
measurement of a high mass will be required to reject these EoS models based on 
mass measurement alone. The maximum spin frequency of $\approx 1250-1500$ Hz can be 
supported by EoS $1-3$ (\S~\ref{Results}). Therefore, spin-down mechanisms, 
such as those due to disk-magnetosphere interaction \citep{Burderietal1999, Ghosh1995}, 
electromagnetic radiation \citep{Ghosh1995} and/or gravitational radiation 
\citep{Bildsten1998}, may be required to explain the the absence of an observed 
spin frequency above 716 Hz. The spin-down due to disk-magnetosphere 
interaction can happen for both accreting ``normal" neutron stars and strange stars,
if the spin-down in the propeller regime is more than the spin-up in the accretion
regime \citep{Burderietal1999}. The electromagnetic radiation and the gravitational
radiation exert negative torques $\propto \nu^3$ and $\propto \nu^5$ respectively 
on the compact star \citep{Ghosh1995,Bildsten1998}.
These two latter mechanisms can also work for strange stars \citep{Ahmedovetal2012,
ander02}.

We note that the maximum angular momentum appears at a lower central density than that
for the maximum mass (Tables~\ref{table_mass_shed_max_mass} and 
\ref{table_mass_shed_max_ang_mom}). This is in agreement with calculations for other
EoS models \citep{Bombacietal2000}.
A comparison between  EoS 1--2 and EoS 3 for the maximum mass or maximum angular momentum
configurations gives some idea about the extent of the dependence of stellar parameter 
values on the EoS model parameter $B_{\rm eff}$ 
(Tables~\ref{table_non_spinning}--\ref{table_mass_shed_max_ang_mom}).
Here we list some of the points.
(1) The central density is significantly lower for the lower effective bag 
constant ($B_{\rm eff}$) value (i.e., EoS 3). This general behaviour is in agreement with the 
scaling law Eq. (\ref{eq:rho6b}) mentioned in \S~\ref{EoS} (see also \S~\ref{appendix}).
(2) A lower effective bag constant value 
can support a larger mass, and the corresponding 
radius is also higher. These are expected from the scaling law Eqs.
(\ref{eq:scal5aa}) and (\ref{eq:scal5bb}), especially for maximum mass configurations
of non-spinning compact stars with EoS model given in Eq. (\ref{eq:eosBag})
(\S~\ref{EoS}; see also \S~\ref{appendix}). We find that, for our EoS 2 and EoS 3, 
the mass ratio and the radius ratio expected from Eqs. (\ref{eq:scal5aa}) and (\ref{eq:scal5bb})
hold within a few percent not only for non-spinning maximum mass configurations 
(Table~\ref{table_non_spinning}),
but also for mass-shed limit maximum mass configurations (Table~\ref{table_mass_shed_max_mass}).
As a result, the stellar compactness, which is the mass-to-radius ratio, 
and hence the surface redshift values, are somewhat similar for all $B_{\rm eff}$ values.
(3) $T/W$ weakly depends on $B_{\rm eff}$, and the value is $\sim 0.21$ for maximum mass
and maximum angular momentum configurations. At this value, the compact star could be
susceptible to triaxial instabilities \citep{Bombacietal2000}.
(4) The oblateness of the compact star, as inferred from $R_{\rm p}/R$ ($\sim 0.53-0.55$), 
is a weak function of $B_{\rm eff}$.
(5) $J$ and $I$ strongly decreases, while $\nu$ significantly increases, with the 
increase of $B_{\rm eff}$. The strong $B_{\rm eff}$-dependence of $I$ is expected from
the scaling law equations \ref{eq:scal5aa} and \ref{eq:scal5bb} (\S~\ref{EoS};
see also \S~\ref{appendix}),
because $I \sim M_{\rm G}R^2$. We verify that $\nu$ roughly scales with $B_{\rm eff}^{1/2}$
(Table~\ref{table_mass_shed_max_mass}),
as expected (see \S~\ref{appendix}). Since $J \propto I\nu $, $J$ is expected to scale 
with $B_{\rm eff}^{-1}$, which we verify (Table~\ref{table_mass_shed_max_mass}).
(6) The two different situations, viz., $r_{\rm ISCO} > R$ and $r_{\rm ISCO} < R$ 
($r_{\rm ISCO}$ is ISCO radius) have
important consequences for observed X-ray features, and hence on the measurements 
of compact star parameters (e.g., \citet{Bhattacharyya2011}).
In the former situation, the length of the gap between $r_{\rm ISCO}$ and $R$ is also very
important for the X-ray emission, because the boundary layer emission to
the accretion disc emission ratio depends on this length (e.g., \citet{Bhattacharyyaetal2000}).
We find that $r_{\rm ISCO}$ is greater than $R$ for the maximum mass and 
maximum angular momentum 
configurations, and the gap-length for EoS 1--2 is somewhat lower than that for 
EoS 3. This is expected from the scaling law equations
\ref{eq:scal5aa} and \ref{eq:scal5bb} (\S~\ref{EoS}; see also \S~\ref{appendix}), 
because $r_{\rm ISCO} \propto M_{\rm G}$.

Let us now examine how $B_{\rm eff}$ can be constrained
from observations, especially when $M_{\rm G}$ and $\nu$ have been measured. 
Comparing EoS 2 ($B^{1/4}_{\rm eff} = 138$ MeV) and EoS 3
($B^{1/4}_{\rm eff} = 125$ MeV), we find significant differences for the
measurable parameters $R/r_{\rm g}$, $R$ and $I$, which are $\approx 17$\%, $\approx 17$\%
and $\approx 31$\% respectively (Table~\ref{table_1.97}).
Note that, since these parameters depend on $a_4$ only very weakly (see row 1 and row 2 of
Table~\ref{table_1.97}),
the relatively small $a_4$ difference between EoS 2 and EoS 3 does not prohibit us to 
study a $B_{\rm eff}$ dependence. 

In order to examine if the above quoted differences are 
in agreement with the scaling laws mentioned in \S~\ref{appendix}, even for a
fast spinning star for which the TOV equations (\ref{eq:TOV1} and \ref{eq:TOV2}) 
are not exactly valid, first we mention the ratios of $R/r_{\rm g}$, $R$ and $I$
for $B^{1/4}_{\rm eff} = 138$ MeV and $B^{1/4}_{\rm eff} = 125$ MeV. These are
$[R/r_{\rm g}]_{\rm EoS 2}/[R/r_{\rm g}]_{\rm EoS 3} = 0.84$,
$R_{\rm EoS 2}/R_{\rm EoS 3} = 0.84$, and 
$I_{\rm EoS 2}/I_{\rm EoS 3} = 0.73$ (Table~\ref{table_1.97}).
Now, since radius is a weak function of central density
for most of the observationally relevant portion of the parameter space 
(say, $M_{\rm G} > 1 M_\odot$; see, for example, Figs.~\ref{mgvslogd1388.ps} and \ref{mgvsre1388.ps}),
we can assume $R \propto B_{\rm eff}^{-1/2}$ from Eq. \ref{eq:scal4b},
and hence we expect $R_{\rm EoS 2}/R_{\rm EoS 3} 
\approx 0.82$, which is close to the above mentioned value. Note that 
$M_{\rm G}$ is constant for all EoS models in Table~\ref{table_1.97}. Therefore,
while comparing parameter values for EoS 2 and EoS 3, we expect
$R/r_{\rm g} \propto R \propto B_{\rm eff}^{-1/2}$ and 
$I \propto R^2 \propto B_{\rm eff}^{-1}$. Hence from \S~\ref{appendix} we expect
$[R/r_{\rm g}]_{\rm EoS 2}/[R/r_{\rm g}]_{\rm EoS 3} = 0.82$
and $I_{\rm EoS 2}/I_{\rm EoS 3} = 0.67$, which are close to the above mentioned values.

So let us now see if some of the above quoted percentage differences of 
$R/r_{\rm g}$, $R$ and $I$ (Table~\ref{table_1.97}) can be observationally measured.
A fortuitous discovery of an atomic spectral line
can constrain $R/r_{\rm g}$ with better than $5$\% accuracy, even when the
star is rapidly spinning making the line broad and skewed \citep{Bhattacharyyaetal2006}.
Other methods to measure $R/r_{\rm g}$ are also available (e.g., \citet{Bhattacharyya2010},
\citet{Bhattacharyyaetal2005}). 
Possible measurements of $R$ for LMXBs and the related difficulties
have been discussed in a number of papers (e.g., 
\citet{Guveretal2012a,Guveretal2012b,Steineretal2010,Steineretal2013,Suleimanovetal2011a,Suleimanovetal2011b,GuillotRutledge2014}).
Modelling of burst oscillations observed with a 
future large area X-ray timing instrument can tightly constrain $R$ of a compact star
in an LMXB \citep{Loetal2013}. The pulsed X-ray emission from radio millisecond
pulsars can also be useful to constrain $R$ 
\citep{Bogdanov2009,Bogdanov2008,Ozeletal2015}. In fact, the upcoming 
{\it NICER} space mission
is expected to measure the $R$ of the nearest and best-studied millisecond 
pulsar PSR J0437-4715 with 5\% accuracy. Besides, a measurement of $I$ of the
binary pulsar J0737-3039A with 10\% accuracy has been talked about \citep{Morrisonetal2004}.
Apart from $R/r_{\rm g}$, $R$ and $I$, [$r_{\rm orb} - R$] of LMXBs, which can be 
inferred from spectral and timing studies of observed X-ray emission (e.g., 
\citet{Bhattacharyyaetal2000}), also has a significant difference ($\approx 98$\%)
between EoS 2 and EoS 3. Therefore, since $M_{\rm G}$ and $\nu$
have been measured for a number of compact stars (see \S~\ref{Introduction}), 
$B_{\rm eff}$ could be constrained from observations, within the ambit of our EoS models.

After discussing how $B_{\rm eff}$ could be constrained,
let us now study the possible effects of the perturbative
QCD corrections term $a_4$ on stellar parameters. 
In order to do this, we compare the stellar properties 
for EoS 1 \& EoS 2 for a constant rest mass sequence $M_{\rm 0} = 2.00 M_{\odot}$ 
(Table~\ref{rest_mass_seq}). This is the sequence along which a non-accreting
compact star evolves. We consider such a sequence for comparison, because 
the $a_4$ value is expected to affect $M_{\rm 0}$ and the total stellar 
binding energy $B$ (see \S~\ref{EoS}). We verify this for three central density 
values (Table~\ref{rest_mass_seq}). 
For all these densities, $B$ for $a_4 = 0.80$ is $\approx 0.12 M_{\odot}$ higher
than $B$ for $a_4 = 0.61$. 

As a result, $a_4$ could have signatures in
evolution of compact star and other system properties, 
within the ambit of our EoS models. 
For example, the evolution of non-accreting 
compact stars happens keeping the $M_{\rm 0}$ value constant, while the other 
parameters evolve depending on the $a_4$ parameter (Table~\ref{rest_mass_seq}).
For accreting compact stars, the increase of $M_{\rm G}$ for a certain amount of
added $M_{\rm 0}$ depends on $B$, and hence on $a_4$. Besides,
orbital period ($P_{\rm orb}$) evolution of LMXBs depends on, among other things,
the fraction of exchanged matter lost from the binary system (see equation 3.14 of
\citet{Bhattacharya1991}). Therefore, 
as an amount of matter $\Delta M_0$ from the companion star
falls on the compact star, the system loses a mass $\Delta M_0 - \Delta M_{\rm G}$.
This is because the transferred matter becomes bound to the compact star, 
and $\Delta M_{\rm G} < \Delta M_0 $. This difference (i.e., the corresponding 
binding energy) is released from the system.
The amount of this lost mass, which affects $P_{\rm orb}$, 
increases with $B$, and hence with $a_4$. Furthermore,
Sudden mass loss can happen when the core of a massive star
collapses into a compact star, or when a ``normal'' neutron star (nucleonic star) changes
into a strange or a hybrid star \citep{r1}, and the star loses gravitational mass as it
becomes more bound. Therefore, $B$ of the final stellar configuration, which will depend
on $a_4$ if the final star contains interacting quark matter, will influence
the stellar parameter values (e.g., $M_{\rm G}$, $\nu$) after collapse. If such a collapsed 
star is in a binary system, then
$B$ (and hence $a_4$) may also significantly affect $P_{\rm orb}$ and the orbital eccentricity
\citep{Flannery1975}. 

Therefore, $a_4$ can be an important ingredient for the computations of stellar evolution
and binary evolution.
Hence the comparison of the results of such computations
with the measured distribution of $M_{\rm G}$, $\nu$, $P_{\rm orb}$ and other 
source parameter values holds the potential to constrain $a_4$. However,
a reliable constraint may be possible, if the systematic uncertainties due to various
unknown source parameters and less understood
processes (e.g., disc-magnetosphere interaction) are sufficiently reduced.

Table~\ref{table_1.97} is additionally useful, because it lists a number of parameter
values of an important pulsar for our EoS models. These values will not only be useful
to constrain EoS models, but also be important to model the accretion and binary
evolution process that created this pulsar. Given the high stellar mass of this source,
such a modelling will be useful to address important problems such as the possibility
of high birth mass of compact stars. 

Table~\ref{table_716} lists a number of parameter values of another pulsar PSR J1748-2446ad.
This is the fastest known pulsar, and hence is of immense importance (\S~\ref{Introduction}). However,
the mass of PSR J1748-2446ad is not known, and hence we compute several stable stellar
configurations for each EoS model, keeping $\nu$ at the observed value. 
These numbers characterize the compact star, and will be useful to study the evolution
that created this pulsar. This study can be important to address problems 
such as why we do not observe a compact star spin frequency higher than a certain value.

\section{Summary}\label{Summary}

Here we summarize the key points of this paper.

(1) We explore the possibility of the existence of strange stars using three
EoS models based on MIT bag model with perturbative corrections due to quark
interactions. We, for the first time, compute the equilibrium sequences 
of fast spinning strange stars for these EoS models.

(2) Our EoS models can support maximum gravitational mass values in the range $\approx 3.0-3.6
M_\odot$, and maximum spin frequencies in the range $\approx 1250-1500$ Hz. Thus
these EoS models are consistent with the maximum measured mass ($\approx 2.0 M_\odot$) and
the highest observed spin frequency ($716$ Hz) of compact stars.

(3) Our EoS models are characterized by two parameters: (a) an effective bag constant
($B_{\rm eff}$), and (b) a perturbative QCD corrections term ($a_4$). We study the effects
of these two parameters on measurable compact star properties.
This could be useful to find possible ways to constrain these fundamental 
quark matter parameters from observations within the ambit of interacting quark matter 
EoS models. 

(4) Effects of $B_{\rm eff}$: we find that a higher stellar
mass is allowed for a lower $B_{\rm eff}$ value.
Furthermore, for a compact star with known gravitational mass and spin frequency,
other measurable parameters, such as stellar radius, radius-to-mass ratio and
moment of inertia, sufficiently increase with the decrease of $B_{\rm eff}$.
These are primarily a consequence of the scaling laws quoted in \S~\ref{appendix}
as discussed in \S~\ref{Discussion}.
Such effects of $B_{\rm eff}$ can be useful to constrain the effective bag constant, 
as mass and spin of compact stars are measurable.

(5) Effects of $a_4$: we find that $a_4$ significantly affects the stellar rest
mass and the total stellar binding energy. Therefore, $a_4$ could 
have signatures in evolutions of both accreting and non-accreting compact 
stars, orbital period evolution of LMXBs, sudden mass loss via collapse, and hence
the observed distribution of stellar mass and spin, orbital period and other source 
parameters.

(6) We compute observationally measurable and other parameter values of two important pulsars:
PSR J1614-2230 and PSR J1748-2446ad for our EoS models. The first one has the highest
precisely measured mass with $< 10$ ms spin period, 
and the second one has the highest measured spin frequency.
Our reported numbers should be useful ingredients for computations of their evolutionary
histories, as well as for constraining EoS models from their future observations.

\section*{Acknowledgements}

We thank an anonymous referee for the constructive comments, which improved the paper.

%%%%%%%%%%%%%%%%%%%%%%%%%%%%%%%%%%%%%%%%%%%%%%%%%%

\clearpage

\begin{table*}
\centering
\caption{Equation of state model parameters used in the present work.}
\begin{tabular}{ccc}
\hline
  EoS & $B_{\rm eff}^{1/4}$~(MeV) & $a_4$ \\
\hline
   1 & 138 & 0.80 \\
   2 & 138 & 0.61 \\
   3 & 125 & 0.50 \\
\hline
\end{tabular}
\label{table_EoS}
\end{table*}

\begin{table*}
\centering
\caption{Stable structure parameters for the nonspinning maximum mass configurations of strange stars
(\S~\ref{Results}).}
\begin{tabular}{ccccccc}
\hline
EoS\footnotemark[1] & $\rho_{\rm c}$\footnotemark[2] & $M_{\rm G}$\footnotemark[3] & $M_{\rm 0}$\footnotemark[4] & $R$\footnotemark[5] & $R/r_{\rm g}$\footnotemark[6] & $r_{\rm orb}$\footnotemark[7] \\
\hline
1 & 17.682 & 2.093 & 2.719 & 11.559 & 3.741 & 18.521 \\
2 & 17.940 & 2.073 & 2.502 & 11.474 & 3.749 & 18.350 \\
3 & 12.553 & 2.479 & 3.090 & 13.736 & 3.753 & 21.893 \\
\hline
\end{tabular}
\begin{flushleft}
$^1$Equation of state models (\S~\ref{EoS} and Fig.~\ref{eos_1.ps}). 
$^2$Central density ($10^{14}$ g cm$^{-3}$). 
$^3$Gravitational mass ($M_{\odot}$). 
$^4$Rest mass ($M_{\odot}$).
$^5$Radius (km).
$^6$Inverse of stellar compactness. Here, $r_{\rm g}$ is the Schwarzschild radius.
$^7$Radius (km) of the innermost stable circular orbit, or the stellar equatorial radius, whichever is bigger.
\end{flushleft}
\label{table_non_spinning}
\end{table*}

\begin{table*}
\centering
\caption{Stable structure parameters for the maximally spinning (i.e., mass-shed limit) 
maximum mass configurations of strange stars (\S~\ref{Results}).}
\begin{tabular}{ccccccccccccccc}
\hline
EoS\footnotemark[1] & $\rho_{\rm c}$\footnotemark[2] & $M_{\rm G}$\footnotemark[3] & $M_{\rm 0}$\footnotemark[4] & $R$\footnotemark[5] & $R/r_{\rm g}$\footnotemark[6] & $R_{\rm p}$\footnotemark[7] & $r_{\rm orb}$\footnotemark[8] & $J$\footnotemark[9] & $\nu$\footnotemark[10] & $I$\footnotemark[11] & $T/W$\footnotemark[12] & $Z_{\rm p}$\footnotemark[13] & $Z_{\rm f}$\footnotemark[14] & $Z_{\rm b}$\footnotemark[15] \\
\hline
%  &  &  &  &  &  &  &  &  &  &  &  &  &  \\
1  & 11.364 & 3.032 & 3.924 & 17.644 & 3.942 & 9.718 & 19.458 & 7.080 & 1412.6 & 7.973 & 0.207 & 0.802 & -0.356 & 2.555 \\
2  & 11.297 & 3.001 & 3.609 & 17.575 & 3.967 & 9.645 & 19.372 & 6.963 & 1415.9 & 7.827 & 0.208 & 0.794 & -0.355 & 2.525 \\
3  & 7.635 & 3.600 & 4.452 & 21.165 & 3.983 & 11.552 & 23.310 & 9.940 & 1170.9 & 13.505 & 0.212 & 0.780 & -0.354 & 2.474 \\
\hline
\end{tabular}
\begin{flushleft}
$^1$Equation of state models (\S~\ref{EoS} and Fig.~\ref{eos_1.ps}). 
$^2$Central density ($10^{14}$ g cm$^{-3}$). 
$^3$Gravitational mass ($M_{\odot}$). 
$^4$Rest mass ($M_{\odot}$).
$^5$Equatorial radius (km).
$^6$Inverse of stellar compactness. Here, $r_{\rm g}$ is the Schwarzschild radius.
$^7$Polar radius (km).
$^8$Radius (km) of the innermost stable circular orbit, or the stellar equatorial radius, whichever is bigger.
$^9$Total angular momentum ($10^{49}$ g cm$^2$ s$^{-1}$).
$^{10}$Spin frequency (Hz).
$^{11}$Moment of inertia ($10^{45}$ g cm$^2$).
$^{12}$Ratio of the total spinning kinetic energy to the total gravitational energy.
$^{13}$Polar redshift.
$^{14}$Forward redshift.
$^{15}$Backward redshift.
\end{flushleft}
\label{table_mass_shed_max_mass}
\end{table*}

\begin{table*}
\centering
\caption{Stable structure parameters for the maximum angular momentum configurations of strange stars (\S~\ref{Results}).}
\begin{tabular}{ccccccccccccccc}
\hline
EoS\footnotemark[1] & $\rho_{\rm c}$ & $M_{\rm G}$ & $M_{\rm 0}$ & $R$ & $R/r_{\rm g}$ & $R_{\rm p}$ & $r_{\rm orb}$ & $J$ & $\nu$ & $I$ & $T/W$ & $Z_{\rm p}$ & $Z_{\rm f}$ & $Z_{\rm b}$ \\
\hline
%  &  &  &  &  &  &  &  &  &  &  &  &  &  \\
1  & 9.784 & 3.020 & 3.892 & 18.081 & 4.056 & 9.670 & 19.922 & 7.173 & 1368.9 & 8.340 & 0.215 & 0.764 & -0.353 & 2.426 \\
2  & 9.372 & 2.970 & 3.550 & 18.110 & 4.130 & 9.567 & 19.939 & 7.010 & 1359.0 & 8.209 & 0.216 & 0.742 & -0.351 & 2.346 \\
3  & 6.891 & 3.583 & 4.415 & 21.519 & 4.068 & 11.504 & 23.684 & 9.979 & 1145.3 & 13.867 & 0.217 & 0.752 & -0.352 & 2.377 \\
\hline
\end{tabular}
\begin{flushleft}
$^1$See Table~\ref{table_mass_shed_max_mass} for meanings of all parameter symbols and units.
\end{flushleft}
\label{table_mass_shed_max_ang_mom}
\end{table*}

\begin{table*}
\centering
\caption{Three stable configurations on the rest mass ($M_{\rm 0} = 2.00 M_{\odot}$) sequence for two
EoS models.}
\begin{tabular}{ccccccccccccccc}
\hline
No.\footnotemark[1] & $\rho_{\rm c}$\footnotemark[2] & EoS & $M_{\rm G}$ & $R$ & $R/r_{\rm g}$ & $R_{\rm p}$ & $r_{\rm orb}$ & $J$ & $\nu$ & $I$ & $T/W$ & $Z_{\rm p}$ & $Z_{\rm f}$ & $Z_{\rm b}$ \\
\hline
%  &  &  &  &  &  &  &  &  &  &  &  &  &  \\
1  & 5.176 & 1 & 1.645 & 14.671 & 6.039 & 8.638 & 16.187 & 2.104 & 1063.3 & 3.149 & 0.169 & 0.325 & -0.179 & 0.909 \\
   &  & 2 & 1.768 & 15.674 & 6.004 & 8.228 & 17.522 & 2.675 & 1112.3 & 3.827 & 0.201 & 0.354 & -0.219 & 1.032 \\
   &  &  &  &  &  &  &  &  &  &  &  &  &  &  \\
2  & 6.190 & 1 & 1.613 & 12.610 & 5.297 & 10.683 & 12.988 & 1.099 & 746.1 & 2.344 & 0.058 & 0.307 & -0.008 & 0.652 \\
   &  & 2 & 1.734 & 13.441 & 5.251 & 10.012 & 14.554 & 1.706 & 949.2 & 2.861 & 0.101 & 0.344 & -0.091 & 0.842 \\
   &  &  &  &  &  &  &  &  &  &  &  &  &  &  \\
3  & 6.866 & 1 & 1.600 & 11.822 & 5.004 & 11.803 & 13.675 & 0.139 & 108.5 & 2.045 & 0.001 & 0.291 & 0.246 & 0.336 \\
   &  & 2 & 1.717 & 12.572 & 4.959 & 10.947 & 13.410 & 1.132 & 719.0 & 2.505 & 0.049 & 0.332 & 0.018 & 0.678 \\
\hline
\end{tabular}
\begin{flushleft}
$^1$Number of a specific ($M_{\rm 0}$, $\rho_{\rm c}$) combination. These numbers are
marked on the $M_{\rm 0} = 2.00 M_{\odot}$ sequence in 
Figs.~\ref{mgvslogd1388.ps}--\ref{omvsj1386.ps}.
$^2$See Table~\ref{table_mass_shed_max_mass} for meanings of all parameter symbols and units.
\end{flushleft}
\label{rest_mass_seq}
\end{table*}

\begin{table*}
\centering
\caption{Stable structure parameters for the gravitational mass 
$M_{\rm G} = 1.97$ M$_\odot$ and spin frequency $\nu = 317.5$ Hz (measured
for PSR J1614-2230) configurations of strange stars (\S~\ref{Results}).}
\begin{tabular}{ccccccccccccc}
\hline
EoS\footnotemark[1] & $\rho_{\rm c}$ & $M_{\rm 0}$ & $R$ & $R/r_{\rm g}$ & $R_{\rm p}$ & $r_{\rm orb}$ & $J$ & $I$ & $T/W$ & $Z_{\rm p}$ & $Z_{\rm f}$ & $Z_{\rm b}$ \\
\hline
%  &  &  &  &  &  &  &  &  &  &  &  &  &  \\
1  & 9.809 & 2.531 & 12.165 & 4.182 & 11.947 & 16.012 & 0.550 & 2.753 & 0.007 & 0.391 & 0.238 & 0.551 \\
2  & 10.309 & 2.355 & 12.062 & 4.147 & 11.854 & 16.038 & 0.540 & 2.707 & 0.007 & 0.396 & 0.244 & 0.556 \\
3  & 4.913 & 2.362 & 14.324 & 4.925 & 13.872 & 15.690 & 0.739 & 3.702 & 0.013 & 0.304 & 0.147 & 0.470 \\
\hline
\end{tabular}
\begin{flushleft}
$^1$See Table~\ref{table_mass_shed_max_mass} for meanings of all parameter symbols and units.
\end{flushleft}
\label{table_1.97}
\end{table*}

\begin{table*}
\centering
\caption{Stable structure parameters for the constant $\nu = 716$ Hz (measured 
for the fastest known pulsar PSR J1748-2446ad) sequence of strange stars (\S~\ref{Results}).}
\begin{tabular}{cccccccccccccc}
\hline
EoS\footnotemark[1] & $\rho_{\rm c}$ & $M_{\rm G}$ & $M_{\rm 0}$ & $R$ & $R/r_{\rm g}$ & $R_{\rm p}$ & $r_{\rm orb}$ & $J$ & $I$ & $T/W$ & $Z_{\rm p}$ & $Z_{\rm f}$ & $Z_{\rm b}$ \\
\hline
%  &  &  &  &  &  &  &  &  &  &  &  &  &  \\
1 & 3.860 & 0.197 & 0.220 & 6.921 & 23.829 & 5.229 & 6.921 & 0.035 & 0.077 & 0.082 & 0.053 & -0.061 & 0.168 \\
  & 5.045 & 1.147 & 1.374 & 11.648 & 6.880 & 9.636 & 11.648 & 0.619 & 1.376 & 0.061 & 0.215 & -0.032 & 0.477 \\
  & 6.230 & 1.607 & 1.989 & 12.521 & 5.278 & 10.777 & 12.812 & 1.037 & 2.305 & 0.052 & 0.304 & 0.004 & 0.633 \\
  & 7.415 & 1.850 & 2.334 & 12.746 & 4.666 & 11.227 & 14.288 & 1.260 & 2.800 & 0.047 & 0.362 & 0.032 & 0.729 \\
  & 8.599 & 1.989 & 2.537 & 12.760 & 4.346 & 11.401 & 15.097 & 1.371 & 3.047 & 0.042 & 0.401 & 0.054 & 0.793 \\
  & 9.784 & 2.072 & 2.663 & 12.691 & 4.149 & 11.473 & 15.596 & 1.421 & 3.157 & 0.039 & 0.430 & 0.071 & 0.838 \\
  & 10.969 & 2.123 & 2.742 & 12.589 & 4.017 & 11.461 & 15.919 & 1.437 & 3.193 & 0.037 & 0.452 & 0.086 & 0.870 \\
  & 12.154 & 2.153 & 2.790 & 12.474 & 3.925 & 11.433 & 16.129 & 1.434 & 3.185 & 0.035 & 0.468 & 0.098 & 0.893 \\
  & 13.338 & 2.171 & 2.819 & 12.356 & 3.855 & 11.377 & 16.260 & 1.419 & 3.152 & 0.033 & 0.482 & 0.108 & 0.910 \\
  & 14.523 & 2.180 & 2.835 & 12.237 & 3.802 & 11.336 & 16.336 & 1.396 & 3.103 & 0.031 & 0.492 & 0.117 & 0.923 \\
  & 16.103\footnotemark[2] & 2.184 & 2.842 & 12.089 & 3.750 & 11.242 & 16.388 & 1.363 & 3.029 & 0.029 & 0.503 & 0.127 & 0.935 \\
  & 17.287 & 2.182 & 2.840 & 11.983 & 3.720 & 11.170 & 16.398 & 1.336 & 2.969 & 0.028 & 0.509 & 0.134 & 0.941 \\
  &  &  &  &  &  &  &  &  &  &  &   &  &  \\
2 & 3.950 & 0.255 & 0.267 & 7.491 & 19.896 & 5.719 & 7.491 & 0.053 & 0.118 & 0.076 & 0.064 & -0.062 & 0.191 \\
  & 5.108 & 1.128 & 1.257 & 11.542 & 6.928 & 9.589 & 11.542 & 0.598 & 1.328 & 0.060 & 0.213 & -0.031 & 0.472 \\
  & 6.265 & 1.574 & 1.809 & 12.404 & 5.338 & 10.678 & 12.404 & 0.996 & 2.212 & 0.052 & 0.300 & 0.004 & 0.622 \\
  & 7.423 & 1.815 & 2.124 & 12.635 & 4.716 & 11.133 & 14.061 & 1.212 & 2.693 & 0.046 & 0.356 & 0.031 & 0.717 \\
  & 8.581 & 1.954 & 2.314 & 12.662 & 4.389 & 11.314 & 14.878 & 1.324 & 2.942 & 0.042 & 0.395 & 0.053 & 0.780 \\
  & 9.739 & 2.040 & 2.434 & 12.601 & 4.185 & 11.393 & 15.386 & 1.378 & 3.060 & 0.039 & 0.424 & 0.070 & 0.825 \\
  & 12.054 & 2.125 & 2.557 & 12.402 & 3.954 & 11.365 & 15.947 & 1.397 & 3.104 & 0.034 & 0.463 & 0.097 & 0.881 \\
  & 14.369 & 2.156 & 2.604 & 12.176 & 3.826 & 11.278 & 16.175 & 1.366 & 3.035 & 0.031 & 0.487 & 0.116 & 0.912 \\
  & 16.299\footnotemark[2] & 2.161 & 2.614 & 11.998 & 3.760 & 11.164 & 16.249 & 1.328 & 2.951 & 0.029 & 0.500 & 0.128 & 0.927 \\
  & 17.456 & 2.160 & 2.612 & 11.896 & 3.731 & 11.096 & 16.258 & 1.302 & 2.894 & 0.028 & 0.506 & 0.135 & 0.933 \\
  &  &  &  &  &  &  &  &  &  &  &  &  &   \\
3 & 2.920 & 0.700 & 0.765 & 12.181 & 11.793 & 8.216 & 12.181 & 0.388 & 0.862 & 0.134 & 0.123 & -0.100 & 0.353 \\
  & 3.913 & 1.764 & 2.057 & 15.246 & 5.856 & 11.798 & 15.575 & 1.647 & 3.660 & 0.089 & 0.281 & -0.067 & 0.664 \\
  & 5.154 & 2.275 & 2.746 & 15.699 & 4.673 & 12.938 & 17.538 & 2.332 & 5.181 & 0.071 & 0.379 & -0.028 & 0.845 \\
  & 6.147 & 2.457 & 3.006 & 15.618 & 4.306 & 13.239 & 18.290 & 2.522 & 5.603 & 0.063 & 0.425 & -0.004 & 0.925 \\
  & 7.139 & 2.552 & 3.149 & 15.443 & 4.099 & 13.348 & 18.725 & 2.578 & 5.727 & 0.057 & 0.457 & 0.015 & 0.975 \\
  & 8.132 & 2.603 & 3.229 & 15.244 & 3.966 & 13.345 & 18.980 & 2.571 & 5.712 & 0.052 & 0.479 & 0.031 & 1.008 \\
  & 9.125 & 2.629 & 3.271 & 15.043 & 3.876 & 13.317 & 19.118 & 2.529 & 5.619 & 0.049 & 0.495 & 0.044 & 1.030 \\
  & 10.118 & 2.639 & 3.290 & 14.852 & 3.812 & 13.239 & 19.189 & 2.474 & 5.497 & 0.046 & 0.507 & 0.054 & 1.045 \\
  & 10.863\footnotemark[2] & 2.640 & 3.294 & 14.714 & 3.774 & 13.188 & 19.204 & 2.427 & 5.393 & 0.044 & 0.515 & 0.061 & 1.052 \\
  & 12.352 & 2.633 & 3.287 & 14.460 & 3.720 & 13.056 & 19.186 & 2.331 & 5.180 & 0.041 & 0.525 & 0.073 & 1.062 \\
\hline
\end{tabular}
\begin{flushleft}
$^1$See Table~\ref{table_mass_shed_max_mass} for meanings of all parameter symbols and units.
$^2$Maximum mass configurations for $\nu = 716$ Hz, and for chosen EoS models.
\end{flushleft}
\label{table_716}
\end{table*}

%%%%%%%%%%%%%%%%% APPENDICES %%%%%%%%%%%%%%%%%%%%%

\appendix

%\section{Appendix}\label{appendix}
\section{}\label{appendix}

It is  known (e.g. \citet{bomb99, Haenseletal2007}) that the mass and the radius 
for non-spinning strange stars, in the case of the EOS given in Eq. (\ref{eq:eosBag}), 
scale with $B_{\rm eff}^{-1/2}$. 
In fact, considering the dimensionless variables: 
\be
{\tilde P} = P/B_{\rm eff}, \qquad\qquad  {\tilde \rho} = c^2\rho/B_{\rm eff} \equiv \varepsilon/B_{\rm eff},
\label{eq:scal1}
\ee            
\be
{\tilde r} = r/r_{\rm o},     \qquad\qquad\qquad  {\tilde m} =  m/m_{\rm o},
\label{eq:scal2}
\ee            
with 
\be 
r_{\rm o} \equiv {{c^2}\over{G^{1/2}B_{\rm eff}^{1/2}}},  \qquad\qquad      
m_{\rm o} \equiv {{c^4}\over{G^{3/2}B_{\rm eff}^{1/2}}}, 
\label{eq:scal3}
\ee
one can easily show that the TOV equations  can be written in the following dimensionless form: 
\be
{{d{\tilde P}}\over{d{\tilde r}}} = 
                         - {{{\tilde m}{\tilde \rho}\over{{\tilde r}^2}} ~
{ {(1 + {{{\tilde P}}\over {{\tilde \rho}}}) 
{(1 + {{4\pi {\tilde r}^3 {\tilde P}}\over { {\tilde m}}} )} \over
{ {(1 - {2 {\tilde m} \over {{\tilde r} }})} }}  }},
\label{eq:TOV1}
\ee
\be 
{d{\tilde m}\over{d{\tilde r}}} = 4 \pi {\tilde r}^2 {\tilde \rho},
\label{eq:TOV2}
\ee
to be solved for any given value of the central density ${\tilde \rho}_c = {\tilde \rho}(0)$
with the boundary conditions ${\tilde m}(0) = 0$ and ${\tilde P}({\tilde R}) = 0$.
Once these dimensionless TOV equations are integrated, the mass and radius 
of the strange star, for an arbitrary value of the constant $B_{\rm eff}$, can be 
obtained from the ``mass'' ${\tilde M}$ and ``radius'' ${\tilde R}$ using 
(\ref{eq:scal1})--(\ref{eq:scal3}): 
\be
M_G(\rho_{\rm c};B_{\rm eff}) = {{c^4}\over{G^{3/2}B_{\rm eff}^{1/2}}} {\tilde M}({\tilde \rho_{\rm c}}),  
\label{eq:scal4a}
\ee
\be
R(\rho_c;B_{\rm eff}) = {{c^2}\over{G^{1/2}B_{\rm eff}^{1/2}}} {\tilde R}({\tilde \rho_{\rm c}}),  
\label{eq:scal4b}
\ee
with the central density $\rho_{\rm c}$ related to the parameter ${\tilde \rho_{\rm c}}$ 
by the second of (\ref{eq:scal1}). 
From equations \ref{eq:scal4a} and \ref{eq:scal4b} one has:
\be
M_G(\rho_{\rm c,1};B_{\rm eff,1}) = \Big( {{B_{\rm eff,2}}\over{B_{\rm eff,1}}}  \Big)^{1/2} 
                                    M_G(\rho_{\rm c,2};B_{\rm eff,2}), 
\label{eq:scal5a}
\ee
\be
R(\rho_{\rm c,1};B_{\rm eff,1}) = \Big( {{B_{\rm eff,2}}\over{B_{\rm eff,1}}}  \Big)^{1/2} R(\rho_{\rm c,2};B_{\rm eff,2}), 
\label{eq:scal5b}
\ee
where $B_{\rm eff,1}$ and $B_{\rm eff,2}$ are two different values of the effective bag constant, and
\be
\rho_{\rm c,1}/\rho_{\rm c,2} = B_{\rm eff,1}/B_{\rm eff,2}.
\label{eq:rho6}
\ee
Equations~(\ref{eq:scal4a})--(\ref{eq:scal5b}) give the scaling law for the 
mass-radius relation. In particular they hold \citep{witt84,hzs86} for the maximum mass configuration. 
Finally, the scaling laws (\ref{eq:scal5a})--(\ref{eq:scal5b}) can be extended to the case 
of spinning configurations. In this case, the stellar structure equations can be written in a dimensionless form  
\citep{Haenseletal2007} if one supplements the dimensionless quantities (\ref{eq:scal1})--(\ref{eq:scal3}) 
with dimensionless angular speeds 
\be 
\tilde{\Omega}  \equiv  {{c}\over{G^{1/2}B_{\rm eff}^{1/2}}}~\Omega,  \qquad\qquad      
\tilde{\omega}  \equiv  {{c}\over{G^{1/2}B_{\rm eff}^{1/2}}}~\omega \,.  
\label{eq:scal7} 
\ee
Thus, the stellar properties for spinning configurations, in the case of the EOS given in Eq. (\ref{eq:eosBag}),  
scale with equations~(\ref{eq:scal5a})--(\ref{eq:rho6}) supplemented with the following scaling law 
for the spin frequency
\be
    \nu_1  =  \Big( {{B_{\rm eff,1}}\over{B_{\rm eff,2}}}  \Big)^{1/2} \, \nu_2 \,. 
\label{eq:scal8}
\ee

%%%%%%%%%%%%%%%%%%%%%%%%%%%%%%%%%%%%%%%%%%%%%%%%%%

%%%%%%%%%%%%%%%%%%%%%%%%%%%%%%%%%%%%%%%%%%%%%%%%%%

% Don't change these lines
\bsp    % typesetting comment
\label{lastpage}
\end{document}